\documentclass{aa}  

\usepackage{graphicx, color}
\usepackage{txfonts}
\usepackage{comment}
\usepackage{booktabs}
\usepackage{placeins}
\usepackage{lscape} 
\usepackage{ragged2e}
\usepackage{orcidlink}
\begin{document}

   \title{Spectropolarimetric characterisation of exoplanet host stars in preparation of the $Ariel$ mission}
\subtitle{II. The magnetised wind environment of TOI-1860, DS Tuc A, and HD 63433 }
   \titlerunning{Spectropolarimetry for the $Ariel$ mission: the magnetic environment of TOI-1860, DS Tuc A, and HD 63433}

   \author{S. Bellotti \inst{1,2}\orcidlink{0000-0002-2558-6920}
          \and
          A. Lavail \inst{2}\orcidlink{0000-0001-8477-5265}
          \and
          D. Evensberget \inst{1,3}\orcidlink{0000-0001-7810-8028}
          \and
          A. A. Vidotto \inst{1}\orcidlink{0000-0001-5371-2675}      
          \and
          C. Danielski \inst{4}\orcidlink{0000-0002-3729-2663}
          \and 
          B. Edwards\inst{5,6}\orcidlink{0000-0002-5494-3237}
          \and
          G. A. J. Hussain \inst{7}\orcidlink{0000-0003-3547-3783}
          \and
          T. L\"uftinger \inst{7}\orcidlink{0009-0000-4946-6942}
          \and
          J. Morin \inst{8}\orcidlink{0000-0002-4996-6901}
          \and
          P. Petit \inst{2}\orcidlink{0000-0001-7624-9222}
          \and
          S. Boro Saikia \inst{9}\orcidlink{0000-0002-3673-3746}
          \and
          G. Micela \inst{10}\orcidlink{0000-0002-9900-4751}
          \and
          A. L\'opez Ariste \inst{2}\orcidlink{0000-0002-1097-8307}
          }
   \authorrunning{Bellotti et al.}
    
   \institute{
            Leiden Observatory, Leiden University,
            PO Box 9513, 2300 RA Leiden, The Netherlands\\
            \email{bellotti@strw.leidenuniv.nl}
        \and
            Institut de Recherche en Astrophysique et Plan\'etologie,
            Universit\'e de Toulouse, CNRS, IRAP/UMR 5277,
            14 avenue Edouard Belin, F-31400, Toulouse, France
        \and Centre for Planetary Habitability (PHAB), Department for Geosciences, University of Oslo, Oslo, Norway
        \and INAF - Osservatorio Astrofisico di Arcetri, Largo E. Fermi 5, 50125, Firenze, Italy
        \and SRON, Netherlands Institute for Space Research, Niels Bohrweg 4, NL-2333 CA, Leiden, The Netherlands
        \and Department of Physics and Astronomy, University College London, Gower Street, WC1E 6BT London, UK
        \and
             Science Engagement and Oversight Office, Directorate of Science, 
             European Space Research and Technology Centre (ESA/ESTEC),
             Keplerlaan 1, 2201 AZ, Noordwijk, The Netherlands
        \and
             Laboratoire Univers et Particules de Montpellier,
             Universit\'e de Montpellier, CNRS,
             F-34095, Montpellier, France
        \and University of Vienna, Department of Astrophysics, Türkenschanzstrasse 17, A-1180 Vienna, Austria
        \and INAF - Osservatorio Astronomico di Palermo, Piazza del Parlamento 1, 90134, Palermo, Italy
             }
   \date{Received ; accepted }

 
  \abstract
   {}   
   {We update the status of the spectropolarimetric campaign dedicated to characterise the magnetic field properties of a sample of known exoplanet-hosting stars included in the current target list of the $Ariel$ mission. The main aims are to inform observing strategies and subsequent analysis of the data of the $Ariel$ mission, and to provide background information on the magnetic properties of the target and their variability on timescales of at least a few years.
   }
   {We analysed spectropolarimetric data collected for 15 G-M type stars with Neo-Narval, HARPSpol, and SPIRou to assess the detectability of the large-scale magnetic field. For three stars we reconstructed the magnetic field topology and its temporal evolution via Zeeman-Doppler imaging (ZDI). Such reconstructions were then used to perform three-dimensional magnetohydrodynamical simulations of the stellar wind and environment impinging on the hosted exoplanets.
   }
   {We detected the magnetic field of six stars. Of these, we performed ZDI reconstructions for the first time of TOI-1860 and DS~Tuc~A, and for the second time of HD~63433, providing temporal information of its large-scale magnetic field. Consistently with previous results on young ($\rm\sim 50-100~Myr$) solar-like stars, the large-scale magnetic field is moderately strong (30-60\,G on average) and complex, with a significant fraction of magnetic energy in the toroidal component and high-order poloidal components. From the simulations of the stellar wind, we found the orbit of TOI-1860~b to be almost completely sub-Alfv\'enic, the orbits of DS~Tuc~A~b and HD~63433~d to be trans-Alfv\'enic, and the orbits of HD~63433~b and c to be super-Alfv\'enic. We obtained marginal detections of the magnetic field for TOI-836 and TOI-2076, and detections for TOI-1136, but the number of observations is not sufficient for magnetic mapping.
   }
   {A magnetic star-planet connection can occur for most of TOI-1860~b's orbit. This can happen more sporadically for DS~Tuc~A~b and HD~63433~c given the lower fraction of their orbit in the sub-Alfv\'enic regime. The orbit of HD~63433~c is nevertheless more sub-Alfv\'enic than previously simulated owing to the temporal evolution of the stellar magnetic field. For HD~63433~b and c, we expect the formation of a bow shock between the stellar wind and the planet despite the evolution of the stellar magnetic field.
   }

   \keywords{Stars: magnetic field --
                Stars: activity --
                Techniques: polarimetric
               }

   \maketitle

%

\section{Introduction}

$Ariel$ is a medium-class (M4) science mission of the European Space Agency planned to be launched in 2029 \citep{Tinetti2018,Tinetti2022}. The aim of $Ariel$ is to perform a comprehensive survey of the chemical composition and structure of exoplanetary atmospheres, in order to understand exoplanets' bulk composition and how planetary systems form and evolve. The mission will target about a thousand transiting planets orbiting different types of host stars, from early A type to late~M \citep{Zingales2018,Edwards2019,Edwards2022}\footnote{The mission candidate sample is available here: \url{https://github.com/arielmission-space/Mission_Candidate_Sample}}.

To perform an informed target selection and optimise the observing strategy of $Ariel$, it is important to characterise exoplanet-hosting stars in a homogeneous manner \citep{Danielski2022,Magrini2022}. This is also crucial to prevent a biased interpretation of the planetary atmosphere and chemistry. Stellar magnetic activity is a known source of uncertainty in differential spectroscopy analyses \citep{Pont2007,Pont2013}, because it can alter the stellar spectrum on the timescale of the transit \citep{Spina2020}. Magnetically induced inhomogeneities on stellar surfaces hinder the precise measurement of both the planetary radius and mass, therefore affecting the atmospheric scale height estimate and the subsequent retrieval \citep{Oshagh2013,Changeat2020,DiMaio2023}. Depending on the coverage and global configuration of these magnetic inhomogeneities, ambiguous atomic and molecular absorption and emission features can occur \citep[e.g.][]{Rackham2018,Rackham2019,Salz2018,Genest2022}. Moreover, the magnetic field of stars governs the amount of short-wavelength radiation \citep[such as extreme ultraviolet and X-rays][]{Gudel2004,Reiners2022} and stellar wind impinging on planets \citep[e.g.][]{Vidotto2015,Garraffo2016,Alvarado-Gomez2022b}, ultimately regulating photochemistry \citep[e.g.][]{Locci2022,Locci2024} and the erosion of planetary atmospheres \citep[e.g.][]{Lammer2003,Lanza2013,McCann2019,Carolan2021b,Hazra2020,Presa2024,vanLooveren2025}. 

In this paper, we update the status of the spectropolarimetric survey aimed at characterising the stellar magnetic field properties and surrounding environment of a representative sample of stars in the current list of potential $Ariel$ targets \citep{Edwards2019,Edwards2022}. The work of \citet{Bellotti2024} introduced such a survey and focussed on the solar-type star HD~63433. With the information on the stellar large-scale magnetic field at hand, stellar wind simulations have shown that young active stars could produce wind conditions that are harsher compared to the present solar neighbourhood \citep[see][for a few examples]{Vidotto2012,Nicholson2016,Folsom2020,Vidotto2023}. Being dictated by the stellar magnetic field, the environment varies over time in correlation with the stellar magnetic cycle \citep[][Smith et al. in review]{McComas2003} and the evolution of the star \citep[e.g.][]{Johnstone2021}. As a result, the region of space in which specific magnetic star-planet interactions take place could be temporarily modified, altering both the nature and intensity of these interactions.

Our survey is divided into three steps \citep[see also][]{Bellotti2024}: 1) a snapshot campaign to assess the detectability of the large-scale magnetic field and optimise further observations, 2) an observing campaign to reconstruct the topology of the surface's stellar magnetic field, and 3) a long-term monitoring to constrain the evolution and variability of the field. The work presented in this paper addresses all of these steps for new targets with respect to \citet{Bellotti2024}. We first describe the snapshot polarimetric observations aimed at determining which stars are suitable for spectropolarimetric follow-up. Then we present the first large-scale magnetic field reconstructions via Zeeman-Doppler imaging (ZDI; \citealt{Semel1989,Donati1997}) for TOI-1860 and DS~Tuc~A. Finally, we provide a new ZDI reconstruction for HD~63433. The large-scale magnetic field topology of this star was previously reconstructed by \citet{Bellotti2024}, and hence we analyse its temporal evolution over a timescale of a year.

This paper is structured as follows. In Sect.~\ref{sec:obs}, we describe the target selection for this study and the spectropolarimetric observations and in Sect.~\ref{sec:detectability} we analyse the detectability of the large-scale magnetic field. In Sect.~\ref{sec:zdi}, we focus on the reconstruction of the large-scale magnetic field of TOI-1860, DS~Tuc~A, and HD\,63433 and in Sect.~\ref{sec:winds} we describe the magnetohydrodynamical simulations of the environment around these stars. We finally discuss our results and draw our conclusions in Sect.\ref{sec:conclusions}.

\section{Observations}\label{sec:obs}

\subsection{Target selection}\label{sec:target_selection}

The construction of the current $Ariel$ target list is described in \citet{Edwards2019}, \citet{Mugnai2020}, and \citet{Edwards2022}. It contains 748 exoplanet systems whose host stars have a spectral type ranging from A ($\rm T_{eff}\sim 10~000~K$) to late M ($\rm T_{eff}\sim 2600~K$). Starting from this list, we applied a series of selection criteria to obtain a sample of stars whose properties are supposedly suitable for magnetometry. The stars should: i) be bright enough in order to carry out observations with a reasonable exposure time and obtain a moderate signal-to-noise ratio (S/N), and ii) exhibit a certain level of magnetic activity as quantified by specific proxies. The latter point would then translate into a potentially detectable large-scale magnetic field.

We applied the following quantitative selection criteria:
\begin{itemize}
    \item Stellar mass $\rm M_*$~$<$~1.2~$\rm M_\odot$. The outer convective envelopes typical of these type of stars are known to produce persistent and intense magnetic fields via dynamo \citep[e.g.][]{Schrijver2000}.
    \item Rotation period $\rm P_\mathrm{rot}<~30~d$, equatorial velocity $\rm v_\mathrm{eq}\sin i>2~km\,s^{-1}$, and chromospheric activity $S$-index~$>$~0.2 (or $\log R'_\mathrm{HK}>-4.8$). These selection limits, corresponding to fast rotation and high magnetic activity, translate into magnetic detection fractions larger than 70\% according to the results of the Bcool snapshot survey \citep{Marsden2014}.
    \item Magnitude in H or V band lower than 9.0. This threshold is applied to limit a polarimetric exposure to a maximum of approximately one hour for spectropolarimeters operating at optical or near-infrared wavelengths.
\end{itemize}

We applied conservative activity thresholds to search for the most active stars in the current list of $Ariel$ targets. Although HD\,63935 and HD\,89345 have $\rm v_\mathrm{eq}\sin i<2~km\,s^{-1}$, we kept them in our list because of their brightness. We also kept HD~260655 albeit its rotation period is 37.5\,d because \citet{Lehmann2024} recently showed that slowly rotating ($\rm P_{rot}\sim100\,d$) M~dwarfs are capable of generating strong magnetic fields. In general, our selection criteria include the mass dependency of the rotation-activity relation \citep[e.g.][]{Pizzolato2003}, but indirectly, since other factors such as brightness dominate our shortlisting. Indeed, most of the M dwarfs in the $Ariel$ Candidate Sample are too faint to be efficiently observed with spectropolarimetry, regardless of their rotation period or activity level. Finally, we did not select stars that have a spectropolarimetric characterisation already, such as GJ~436 \citep{Bellotti2023a,Vidotto2023} and AU~Mic \citep{Klein2021,Donati2023} for instance. 

The stellar sample analysed in this work contains 15 objects and its properties are listed in Table~\ref{tab:prelim_mag_info}. The fraction of low-mass, highly-to-moderately active stars in the Ariel Candidate Sample that are shortlisted with our selection criteria excluding the magnitude criterion is $\sim\,30\%$. Our 15 stars are therefore representative of the bright end of this specific parameter space. We note that for $\log R'_\mathrm{HK}$ we are currently using literature values, but that in the future publications we will use homogeneous data that are currently being prepared by Wizani et al, in prep. We also note that the list of targets for $Ariel$ is being optimised over time, hence some of the targets selected for this spectropolarimetric campaign may have been removed from the mission candidate sample, but they remain interesting targets (e.g. for other follow-up efforts than $Ariel$).

\begin{table*}[ht]
\caption{Properties of the selected stars.}
\label{tab:prelim_mag_info}     
\centering                       
\begin{tabular}{l c c c c c c c c c c c}      
\toprule     
Star & RA & DEC & SpT & Mass & H & V & $\rm P_{rot}$ & $\rm v_{eq}\sin(i)$ & $S$-index & $\rm \log R'_\mathrm{HK}$ & $\rm N_{pl}$\\
 & & & & [M$_\odot$] & & & [d] & [km s$^{-1}$] & & &\\
\midrule
HD\,63433    & 07 49 55.06 & +27 21 47.46   & G5 & 0.99 & 5.36 & 6.92  & 6.5$^a$   & 7.3$^a$  & $\cdots$ & $-4.36^b$ & 3    \\
HD\,63935    & 07 51 41.99 & +09 23 09.79   & G5 & 0.97 & 6.96 & 8.58  & $\cdots$  & $0.3^c$ & $\cdots$ & $\cdots$ & 2\\ 
HD\,89345    & 10 18 41.06 & +10 07 44.50   & G5 & 1.18 & 7.77 & 9.38  & $\cdots$  & $0.1^c$  & $\cdots$ & $\cdots$ &1\\
HD\,152843   & 16 55 08.36 & +20 29 28.79   & G0 & 1.06 & 7.66 & 8.86  & $\cdots$  & $7.5^c$  & $\cdots$ & $\cdots$ &2\\ 
HD\,158259   & 17 25 24.10 & +52 47 26.47   & G0 & 1.03 & 5.03 & 6.48  & 20.0$^d$  & $1.6^c$  & 0.19$^e$ & $-4.75^e$ & 5\\
HD\,260655   & 06 37 10.80 & +17 33 53.33   & M0 & 0.44 & 6.03 & 9.59  & 37.5$^f$  & $<2.0^f$ & 0.92$^e$ & $-4.84^e$ & 2\\
HAT-P-22     & 10 22 43.59 & +50 07 42.06   & G5 & 1.13 & 7.94 & 9.76  & 28.7$^g$  & 1.65$^g$ & 0.15$^e$ & $-5.09^e$ & 1\\
DS\,Tuc\,A   & 23 39 39.48 & $-$69 11 44.71 & G5 & 1.01 & 6.76 & 8.23  & 2.85$^h$  & 17.8$^h$ & 0.88$^e$ & $-4.06^e$ & 1\\
Kepler-444   & 19 19 00.55 & +41 38 04.58   & K0 & 0.67 & 6.77 & 8.87  & 17.2$^i$  & $\cdots$ & $\cdots$ & $\cdots$ & 5\\
K2-116       & 22 24 36.38 & $-$11 34 43.30 & K7 & 0.69 & 8.02 & 10.80 & 29.8$^j$  & $\cdots$ & $\cdots$ & $\cdots$ & 1\\
TOI-836      & 15 00 19.40 & $-$24 27 14.69 & K4 & 0.68 & 6.98 & 9.92  & 21.99$^k$ & 1.9$^k$  & $1.31^e$ & $-4.44^e$ & 2\\
TOI-1136     & 12 48 44.37 & +64 51 19.15   & G5 & 1.02 & 8.09 & 9.53  & 8.19$^l$  & 6.7$^m$  & 0.32$^m$ & $-4.49^m$ & 6\\
TOI-1860     & 15 05 49.90 & +64 02 49.94   & G5 & 0.99 & 6.88 & 8.40  & 4.43$^n$  & 10.4$^n$ & $\cdots$ & $-4.25^n$ & 1\\
TOI-2076     & 14 29 34.24 & +39 47 25.54   & K1 & 0.82 & 7.19 & 9.14  & 6.84$^o$  & 5.7$^p$ & $\cdots$ & $-4.37^o$ & 3 \\ 
Wolf~503    & 13 47 23.44 & $-$06 08 12.73 & K3.5 & 0.63 & 7.77 & 10.30 & $\cdots$  & $2.8^c$ & $\cdots$ & $\cdots$ & 1 \\
\bottomrule
\end{tabular}
\tablefoot{The columns are: identifier, right ascension and declination (J2000), spectral type, stellar mass, H band magnitude, V band magnitude, rotation period, projected equatorial velocity, $S$-index, chromospheric activity index, and number of exoplanets. The information is primarily extracted from the $Ariel$ stellar catalogue \citep{Danielski2022} available at \url{https://sites.google.com/inaf.it/arielstellarcatalogue/}. References: $a$. \citet{Mann2020}  $b$. \citet{Marsden2014}, $c$. \citet{Tsantaki2025}, $d$. \citet{Hara2020}, $e$. \citet{BoroSaikia2018}, $f$. \citet{Luque2022}, $g$. \citet{Mancini2018}, $h$. \citet{Newton2019}, $i$. \citet{Angus2018}, but \citet{Mazeh2015} reported a rotation period of 49.4\,d instead, $j$. \citet{Reinhold2020}, $k$. \citet{Hawthorn2023}, $l$. \citet{Canto2020}, $m$. \citet{Dai2023}, $n$. \citet{Giacalone2022}, $o$. \citet{Osborn2022} and \citet{Frazier2023} reported a rotation period of 7.12\,d, and $p$. \citet{Frazier2023}.}
\end{table*}

\subsection{Instruments}

We analysed spectropolarimetric observations collected with Neo-Narval, HARPSpol, and SPIRou. The journal of the observations can be found in Table~\ref{tab:log}, excluding the 2023 observations of HD~63433 published in \citet{Bellotti2024}.

Neo-Narval\footnote{\url{https://www.news.obs-mip.fr/neo-narval-pic-du-midi/}} is the upgraded version of Narval \citep{LopezAriste2022} mounted on the 2\,m T\'elescope Bernard Lyot (TBL) at the Pic du Midi Observatory in France \citep{Donati2003}. The upgrade occurred in 2019, and kept the main performances of Narval: a spectral coverage from 380 to 1050 nm, and a median spectral resolving power of $\sim 65~000$ after data reduction \citep{LopezAriste2022}.

HARPSpol\footnote{\url{https://www.eso.org/sci/facilities/lasilla/instruments/harps.html}} \citep{2011ASPC..437..237S, 2011Msngr.143....7P} is the spectropolarimeter for the HARPS spectrograph \citep{2003Msngr.114...20M} mounted at the Cassegrain focus of the ESO 3.6~m telescope at La Silla observatory, Chile. HARPSpol observations cover the wavelength range between 380 and and 691~nm with a 8~nm gap at 529~nm separating the red and blue detectors. The resolving power is $ R \sim 110~000$. The data reduction was carried out with the {\tt PyReduce} package\footnote{\url{https://github.com/AWehrhahn/PyReduce}} \citep{2021A&A...646A..32P}, the updated implementation in {\tt python} of the versatile {\tt REDUCE} package \citep{2002A&A...385.1095P}. The reduction with {\tt PyReduce} is run with a series of standard steps, similar to {\tt REDUCE} as described by \citet{2013A&A...558A...8R}. For more information on the steps applied to our observations see \citet{Bellotti2024}.

The SpectroPolarim\`etre InfraRouge (SPIRou\footnote{\url{https://www.cfht.hawaii.edu/Instruments/SPIRou/}}) is the near-infrared spectropolarimeter mounted at Cassegrain focus on the 3.6\,m CFHT atop Maunakea, Hawaii \citep{Donati2020}. The instrument allows linear and circular polarisation observations at a spectral resolving power of $R \sim 70\,000$ for a wavelength coverage between 960 to 2500~nm ($YJHK$ bands). Optimal extraction of SPIRou spectra was carried out with {\it A PipelinE to Reduce Observations} (\texttt{APERO} v0.6.132), a fully automatic reduction package installed at CFHT \citep{Cook2022}. 

Our observations were carried out in circular polarisation mode. The output consists of unpolarised (Stokes~$I$), circularly polarised (Stokes~$V$) and null (Stokes~$N$) high-resolution spectra. The Stokes~$N$ spectrum is practical to check the presence of spurious polarisation signatures or data reduction issues \citep[see][for more details]{Donati1997,Bagnulo2009,Tessore2017}.

\section{Magnetic field detectability}\label{sec:detectability}
\subsection{Least-squares deconvolution}

For the first part of our campaign dedicated to the detectability of the large-scale magnetic field, we requested a handful of observations per star at separated days throughout the observing semester. We applied least-squares deconvolution \citep[LSD;][]{Donati1997,Kochukhov2010a} to the collected unpolarised, circularly polarised and null spectra. We used the \texttt{pylsd} python code\footnote{available at \url{https://github.com/folsomcp/LSDpy}} which is part of the \texttt{specpolflow} package \citep{Folsom2025} to deconvolve the spectra with a synthetic atomic line list. Together with wavelength, a line list typically contains the associated atomic number, depth, excitation potential, and effective Land\'e factor (indicated as $\rm g_{eff}$), which encapsulates the sensitivity to Zeeman effect. The deconvolution results in an single, high-signal-to-noise ratio (S/N) kernel (for Stokes $I$, $V$, and $N$) summarising the properties of hundreds or thousands of spectral lines. 

We generated synthetic line lists via the Vienna Atomic Line Database\footnote{\url{http://vald.astro.uu.se/}} \citep[VALD,][]{Ryabchikova2015}, selecting models whose effective temperature and surface gravity are close to the values reported in the $Ariel$ stellar group catalogue for each target. For observations at optical wavelengths with Neo-Narval and HARPSpol, we chose atomic lines between 350 and 1097\,nm, with known $\rm g_{eff}$, and whose depth relative to the continuum is larger than 40\% \citep[following][]{Moutou2007}. We adopted an LSD normalisation wavelength of 700\,nm and a normalisation $\rm g_{eff}$ of 1.2. For observations at near-infrared wavelengths with SPIRou, we selected atomic lines between 950 and 2600\,nm, with known $\rm g_{eff}$, and with a relative depth larger than 3\% \citep[see e.g.][]{Bellotti2023b}. In this case we adopted a normalisation wavelength of 1700\,nm and a normalisation Land\'e factor of 1.2.

The properties of the synthetic line lists used in this work are summarised in Table~\ref{tab:lsd_masks}, in which we also report the total number of spectropolarimetric observations. To quantify whether a circularly polarised Zeeman signature is detected, and thus the large-scale magnetic field, we computed the false-alarm probability \citep[FAP; see][for more details]{Donati1997}. A successful detection corresponds to $\rm FAP<10^{-4}$, and a marginal detection to $\rm FAP=10^{-2}-10^{-4}$. The same metric is applied to Stokes~$N$ in order to check for spurious polarisation signatures. The results are listed in Table~\ref{tab:log}.

\setlength{\tabcolsep}{3pt}
\begin{table}[!t]
\caption{Properties of the synthetic line lists.} 
\label{tab:lsd_masks}     
\centering                       
\begin{tabular}{l c c c c c c}    
\toprule
Name & Instrument & N$_\mathrm{obs}$ & Mask & $\langle\lambda\rangle$ & $\rm\langle g_{eff}\rangle$ & N$_\mathrm{lines}$ \\
& & & & [nm] & & \\
\midrule
HD~63433   & Neo-Narval & 9  & 5500\_4.5 & 447  & 1.23 & 3826 \\
HD~63935   & Neo-Narval & 4  & 5500\_4.5 & 447  & 1.23 & 3827 \\ 
HD~63935   & HARPSpol   & 4  & 5500\_4.5 & 447  & 1.23 & 3682 \\ 
HD~89345   & Neo-Narval & 5  & 5500\_4.0 & 440  & 1.21 & 5167 \\ 
HD~152843  & Neo-Narval & 3  & 6250\_4.0 & 430  & 1.21 & 3401 \\ 
HD~158259  & Neo-Narval & 3  & 5750\_4.5 & 440  & 1.23 & 3309 \\ 
HD~260655  & SPIRou     & 1  & 4000\_5.0 & 1680 & 1.25 & 2432 \\ 
HAT-P-22   & Neo-Narval & 2  & 5500\_4.5 & 447  & 1.23 & 3826 \\ 
DS~Tuc~A   & HARPSpol   & 10 & 5500\_4.5 & 447  & 1.23 & 3702 \\
Kepler-444 & Neo-Narval & 4  & 5000\_4.5 & 443  & 1.22 & 6046 \\ 
K2-116     & SPIRou     & 3  & 4500\_4.5 & 1700 & 1.24 & 3868 \\ 
TOI-836    & SPIRou     & 1  & 4500\_4.5 & 1700 & 1.24 & 3764 \\ 
TOI-1136   & Neo-Narval & 6  & 5750\_4.5 & 440  & 1.23 & 3311 \\ 
TOI-1860   & Neo-Narval & 20 & 5750\_4.5 & 440  & 1.23 & 3313 \\ 
TOI-2076   & Neo-Narval & 6  & 5000\_4.5 & 443  & 1.22 & 6046 \\ 
Wolf~503   & SPIRou     & 3  & 4500\_4.5 & 1700 & 1.24 & 3905 \\ 
\bottomrule 
\end{tabular}
\tablefoot{The columns indicate: identifier of the star, instrument used for the observations, number of observations, properties of the line list in the form $\rm T_{eff}$[K]\_logarithm surface gravity[cm\,s$^{-1}$], number of lines used in LSD, average wavelength of the mask, and average effective Land\'e factor of the mask. The lower number of lines used for HARPSpol stems from the narrower wavelength coverage compared to Neo-Narval. For HD~63433, the values refer to the observations taken in 2024 (see also Table~\ref{tab:log}). }
\end{table}
\setlength{\tabcolsep}{6pt}

Zeeman signatures in circular polarisation are detected for TOI-1136, TOI-1860, and DS~Tuc~A, as well as for the newest observations of HD~63433. As a result, these stars are amenable to magnetic field characterisation. For TOI-836 we obtained one marginal detection and for TOI-2076 we obtained two while the remaining four observations are non-detections. While for TOI-836, TOI-1136, and TOI-2076 we require additional observations for producing a ZDI map, for TOI-1860, HD~63433, and DS~Tuc~A we collected a number of observations that sample various longitudes and are sufficient for the ZDI reconstruction (see Sect.~\ref{sec:zdi}).

We did not detect circularly polarised Zeeman signatures in eight stars over multiple snapshot observations. This is not completely surprising as exoplanets are more easily detected around magnetically inactive stars, that is with a weak large-scale magnetic field. Another possibility is that the large-scale configuration of these stars is such that the magnetic polarity cancellation is substantial, although distributing multiple observations across the semester should allow for some detections owing to stellar rotation and magnetic field variability. This aspect makes the substantial polarity cancellation scenario less likely. The number of non-detections of our stars are overall in agreement with the statistics of the BCool campaign \citep{Marsden2014}, for which a moderate fraction of the large-scale magnetic field of stars with surface temperature between 5250\,K and 6000\,K (that is, spectral types G5 to G0) is not detected. This is also noted considering the stellar rotation period, since most of our non-detections are for slowly rotating stars with $\rm P_{rot}>20\,d$.

\subsection{Longitudinal magnetic field}\label{sec:longitudinal}

The longitudinal magnetic field is the net, line-of-sight-projected component of the large-scale magnetic field \citep[e.g.][]{Donati1997}. We used the general formula as in \citet{Cotton2019}, 
\begin{equation}
\rm B_l = \frac{h}{\mu_B \lambda_0 g_{eff}}\frac{\int vV(v)dv}{\int(I_c-I(v))dv} \,,
\label{eq:Bl}
\end{equation}
where $\lambda_0$ and $\rm g_{eff}$ are the normalisation wavelength and Land\'e factor of the LSD profiles, $\rm I_c$ is the continuum level, $\rm v$ is the radial velocity associated to a point in the spectral line profile in the star's rest frame, $h$ is the Planck's constant and $\rm \mu_B$ is the Bohr magneton. To express it as in \citet{Rees1979} and \citet{Donati1997}, one can use $hc/\mu_B=0.0214$\,Tm, where $c$ is the speed of light in m\,s$^{-1}$.

We computed the longitudinal magnetic field (B$_l$) for TOI-836, TOI-1136, TOI-1860, TOI-2076, HD~63433, and DS~Tuc~A using the continuum-normalised Stokes profiles. In practice, we fitted a linear model to the region outside the Stokes~$I$ LSD profile, to include residuals of continuum normalisation at the level of the spectra, and we re-scaled the Stokes~$V$ LSD profiles by the same fit. For TOI-1136 and TOI-2076, we used only the observations in which Stokes~$N$ detection is at most marginal. For the stars without detections, we computed $3\sigma$ upper limits. The results are given in Table~\ref{tab:log}.

For TOI-1136, the integral range was set to $\rm \pm15\,km\,s^{-1}$. For the two observations when Stokes~$V$ was marginally and completely detected and Stokes~$N$ showed only a marginal detection (18 and 19 March 2024), we obtained $\rm 25\pm44\,G$ and $\rm 131\pm60\,G$, where the error bar indicates the 1$\sigma$ formal uncertainty. We caution that the presence of a marginal detection in Stokes~$N$ can alter the polarisation signature in Stokes~$V$, ultimately affecting the value of $\rm B_\ell$. For TOI-2076 we integrated Eq.~\ref{eq:Bl} between $\rm \pm20\,km\,s^{-1}$ and obtained $\rm -75\pm17\,G$ and $\rm 69\pm34\,G$.

For TOI-1860, the integral range was set to $\rm \pm25\,km\,s^{-1}$. For the observations in 2024 with a detection, we found values between $\rm -5\,G$ and 19\,G and for the 2025 observations we found values between $\rm -21\,G$ and 20\,G. For DS~Tuc~A, we used a range of $\rm \pm35\,km\,s^{-1}$ and obtained values between $\rm -36\,G$ and $\rm -4\,G$, with an average of $\rm -19\,G$. For HD~63433, we applied Eq.~\ref{eq:Bl} between $\rm \pm20\,km\,s^{-1}$ from the line centre at $\rm -15.9\,km\,s^{-1}$. We found values between $\rm -6\,G$ and $\rm 5\,G$, with an average of $\rm -1.0\,G$, similar to the measurements on the 2023 data \citep{Bellotti2024}. 

\section{Large-scale magnetic field reconstruction}\label{sec:zdi}

The second and third parts of our spectropolarimetric campaign are dedicated to reconstructing the stellar large-scale magnetic field for the first time or across different epochs. So far, we have performed enough monitoring to reconstruct a ZDI map for TOI-1860, DS~Tuc~A, and HD~63433. For TOI-1860, we only used the June and July 2025 data set, since the August 2025 data yielded non-detections and the 2024 data set did not have enough observations. The ZDI algorithm inverts a time series of Stokes~$V$ LSD profiles into a magnetic field map, by iteratively synthesising and adjusting model Stokes~$V$ profiles, until a maximum-entropy solution at a fixed reduced $\chi^2$ is achieved \citep[for more information, see][]{Skilling1984,DonatiBrown1997,Folsom2018}. The field is formally described as the sum of a poloidal and toroidal component, which are both expressed via spherical harmonic decomposition \citep{Donati2006,Lehmann2022}. The odd and even spherical harmonics modes are weighted equally in the reconstruction. We employed the \texttt{zdipy} code described in \citet{Folsom2018}. 

We adopted the weak-field approximation because local field strengths are lower than 160\,G for our targets (see Table~\ref{tab:zdi_output}), that is, within the range of applicability of the approximation \citep[typically less 1~kG][]{Kochukhov2010a}. We note that magnetic field strengths at unresolved spatial scales are most likely larger than $\rm1\,kG$ as corroborated by Zeeman broadening and intensification measurements \citep[see][for some recent results]{Kochukhov2020,Hahlin2023}. Assuming weak-field approximation, Stokes~$V$ is proportional to the derivative of Stokes~$I$ with respect to wavelength \citep[e.g.][]{Landi1992}. The unpolarised line profiles in each cell of the stellar grid are modelled with a Voigt kernel, parametrised by the line depth ($d$), Gaussian width ($w_G$), and Lorentzian width ($w_L$). The optimal values of these three parameters are obtained with a $\chi^2_r$ minimisation between the median of the observed Stokes~$I$ LSD profiles and its model over a grid of ($d$, $w_G$, $w_L$) values. We provide the optimal values for each star below.

The stellar input parameters of ZDI are the rotation period $\rm P_{rot}$, the projected equatorial velocity $\rm v_{eq}\sin(i)$, and the inclination $i$ (that is, the viewing angle). Furthermore, the \textsc{zdipy} code includes surface shear as a function of colatitude ($\theta$) expressed in the form
\begin{equation}\label{eq:diff_rot}
\rm \Omega(\theta) = \Omega_{eq} - d\Omega\sin^2(\theta),
\end{equation}
where $\rm \Omega_{eq}=2\pi/P_{rot}$ is the rotational frequency at equator and $d\Omega$ is the differential rotation rate in $\rm rad\,d^{-1}$. The variable $d\Omega$ is also an input parameter of ZDI.

\begin{figure}[t]
    \includegraphics[width=\columnwidth]{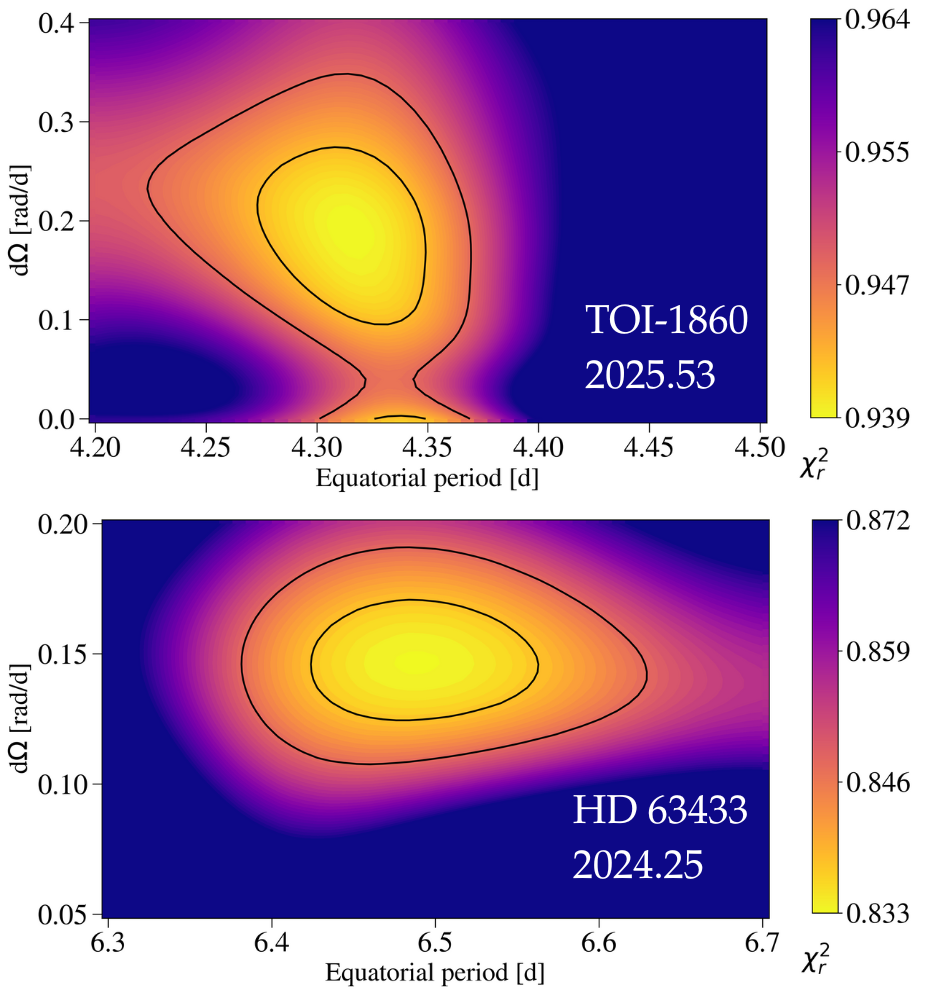}
    \caption{Joint search of differential rotation and equatorial rotation period for TOI-1860 (top) and HD~63433 (bottom). The panels illustrate the $\chi^2_r$ landscape over a grid of (P$_\mathrm{rot,eq}$,$d\Omega$) pairs, with the $1\sigma$ and $3\sigma$ contours. The best values are obtained by fitting a 2D paraboloid around the minimum, while their error bars are estimated from the projection of the $1\sigma$ contour on the respective axis \citep{Press1992}.}
    \label{fig:domega}
\end{figure}

The input $\rm v_{eq}\sin(i)$ used for TOI-1860, DS~Tuc~A, and HD~63433 is listed in Table~\ref{tab:prelim_mag_info}. The stellar inclination was estimated comparing the stellar radius provided in the stellar working group catalogue with the projected radius $R\sin i=\mathrm{P}_\mathrm{rot}v_\mathrm{eq}\sin i/50.59$, where $R\sin(i)$ is measured in solar radii, P$_\mathrm{rot}$ in d, and $v_\mathrm{eq}\sin i$ in km\,s$^{-1}$. We estimated $i\simeq90^{\circ}$ for DS~Tuc~A and TOI-1860, so we adopted a value of $i=80^{\circ}$ in ZDI to prevent mirroring effects between the northern and southern hemispheres that a value of $90^{\circ}$ would incur. For HD~63433, we used $i=70^{\circ}$ as constrained by \citet{Mann2020}. 

For the latitudinal differential rotation, we performed the parameter optimisation outlined in \citet{Donati2000} and \citet{Petit2002}. We generated a grid of ($\rm P_{rot}$, $d\Omega$) pairs, reconstructed a ZDI map for each of the pairs, and searched for the values that minimised the $\chi^2_r$ distribution between observations and synthetic LSD profiles, at a fixed entropy level. The best parameters are measured by fitting a 2D paraboloid to the $\chi^2_r$ distribution, and the error bars are obtained from a variation of $\Delta\chi^2_r = 1$ away from the minimum \citep{Press1992,Petit2002}. If no differential rotation is constraint, we assumed solid body rotation with $\rm P_{rot}$ from Table~\ref{tab:prelim_mag_info}.

In general, a faster rotation (that is, larger $\rm v_{eq}\sin(i)$) translates into a decreased polarity cancellation because the Zeeman signatures are more separated in radial velocity space \citep[see][for example]{Hussain2009}. For this reason, we set the maximum degree of spherical harmonic coefficients $\ell_\mathrm{max}$ to $10$ for every ZDI reconstruction. We note that a lower value could have been used, since most of the magnetic energy is stored in the $\ell<6$ degrees. Finally, we fixed the $V$-band limb darkening coefficient to 0.6964 \citep{Claret2011}. 

For each of the three stars, the observations were phased with the following ephemeris
\begin{equation}
    \rm HJD = HJD_0 + P_{rot} n_{cyc},\\
    \label{eq:ephemeris}
\end{equation}
where $\rm P_{rot}$ is the stellar rotation period and $\rm n_{cyc}$ is the rotation cycle. The reference heliocentric Julian date $\rm HJD_0$ is the first observation or the median observation, depending on whether solid body rotation or differential rotation was assumed.

\setlength{\tabcolsep}{4pt}
\begin{table*}[!t]
\caption{Properties of the large-scale magnetic field and wind of the studied stars.} 
\label{tab:zdi_output}     
\centering                       
\begin{tabular}{l c c c c c c c c c c c c c c}      
\toprule
Star & $\rm \langle|B_V|\rangle$   & $\rm |B_{V,max}|$ & $\rm \langle B_V^2\rangle$ & $\rm f_{pol}$ & $\rm f_{tor}$  & $\rm f_{dip}$   & $\rm f_{quad}$  & $\rm f_{oct}$ & $\rm f_{\ell=4}$  & $\rm f_{axi}$ & $\rm f_{axi,pol}$ & $\rm f_{axi,tor}$ & $\rm \dot{M}$ & $\rm \dot{J}$\\
& [G] & [G] & [$\rm G^2$] & [\%] & [\%] & [\%] & [\%] & [\%] & [\%] & [\%] & [\%] & [\%] & [$\rm M_\odot/yr$] & [erg]\\
\midrule
TOI-1860 & 37 & 103 & $1.8\times10^3$ & 28 & 72 & 28 & 13 & 26 & 20 & 73 & 11 & 98 & $1.5\times10^{-13}$ & $1.4\times10^{31}$\\
DS~Tuc~A & 64 & 160 & $4.9\times10^3$ & 64 & 36 & 67 & 12 & 9 & 5 & 76 & 63 & 98 & $4.4\times10^{-13}$ & $1.2\times10^{32}$ \\
HD~63433 (2023) & 24 & 54 & $0.7\times10^3$ & 46 & 54 & 30 & 25 & 15 & 10 & 66 & 38 & 90 & $1.1\times10^{-13}$ & $4.9\times10^{30}$\\
HD~63433 (2024) & 30 & 71 & $1.1\times10^3$ & 47 & 53 & 7 & 23 & 20 & 36 & 51 & 7 & 91 & $1.4\times10^{-13}$ & $4.5\times10^{30}$\\
\bottomrule                                
\end{tabular}
\tablefoot{The following quantities are listed: star name, mean unsigned large-scale magnetic strength, maximum unsigned large-scale magnetic strength, total reconstructed magnetic energy, poloidal and toroidal magnetic energies as a fraction of the total energy, dipolar, quadrupolar, and octupolar magnetic energy as a fraction of the poloidal energy, axisymmetric magnetic energy as a fraction of the total energy, poloidal axisymmetric energy as a fraction of the poloidal energy, toroidal axisymmetric energy as a fraction of the toroidal energy, mass loss rate, and angular momentum loss rate. The first row of HD~63433 refers to the 2023 ZDI reconstruction \citep{Bellotti2024} while the second row to the 2024 reconstruction presented here.}
\end{table*}
\setlength{\tabcolsep}{6pt}

\subsection{TOI-1860}

As reported in Table~\ref{tab:prelim_mag_info}, TOI-1860 has a rotation period of 4.43\,d and $\rm v_{eq}\sin(i)=10.4~km\,s^{-1}$ \citep{Giacalone2022}. The optimised Voigt kernel parameters are $d=1.95$, $w_G=1.12\rm\,km\,s^{-1}$, and $w_L=3.0\rm\,km\,s^{-1}$ and the assumed inclination is $80^\circ$. Our ten Neo-Narval observations span 23\,d (or 5.2 rotational cycles), so a surface shear would have had the time to distort the LSD profiles sufficiently to be detected. The result of the differential rotation search is shown in Fig.~\ref{fig:domega}. The analysis yielded $\rm P_{rot}=4.315\pm0.035\,d$ and $\rm d\Omega=0.189\pm0.089~rad~d^{-1}$, which implies a rotation period at the pole of $\rm P_{rot}=4.959\pm0.351\,d$ (see Eq.~\ref{eq:diff_rot}) and an equator-pole lap time of 33\,d. The value of $\rm d\Omega$ is consistent with typical latitudinal differential rotation rates of young solar-like stars \citep[see][for a recent example]{Bellotti2025}. Although the $\rm \chi^2_r$ landscape features a minimum, the lower boundary of the grid features a decrease in $\rm \chi^2_r$ making the overall landscape shape irregular. This is reflected in a large error bar of the differential rotation rate. Moreover, \citet{Petit2002} showed that when estimating $\rm d\Omega$ with fewer phases than about 15, rotational phase gaps tend to generate biases that are not included in the statistical error bar.

The model Stokes~$V$ profiles were fit down to $\rm \chi^2_r=0.98$ (see Fig.~\ref{fig:stokesV}), from an initial value of 2.05 which corresponds to a featureless magnetic map. The target $\chi^2_r$ is determined by running ZDI over a grid of $\chi_r^2$ values with all other parameters fixed, each time recording the entropy at convergence, and by measuring the maximum of the change rate in the entropy \citep[see][for more details]{Alvarado-Gomez2015}.

The ZDI magnetic map is illustrated in Fig.~\ref{fig:zdi_maps} and its properties are listed in Table~\ref{tab:zdi_output}. The unsigned, mean magnetic field strength is $\rm B_{mean}=37\,G$, and the topology is predominantly toroidal, as the corresponding component accounts for 72\% of the total magnetic energy. Of the poloidal component, the dipolar, quadrupolar, and octupolar modes store 28\%, 13\%, and 26\% of the energy. The large-scale toroidal component is mostly axisymmetric (98\%) while the poloidal component is non-axisymmetric (11\%).

\subsection{DS Tuc A}

For DS~Tuc~A, the optimal Voigt kernel parameters are: $d=0.52$, $w_G=1.75\rm\,km\,s^{-1}$, and $w_L=4.0\rm\,km\,s^{-1}$. We set the rotation period to 2.85\,d, the $\rm v_{eq}\sin(i)$ to 17.8~km~s$^{-1}$ \citep{Newton2019}, and the inclination to $80^\circ$. The search for latitudinal differential rotation was inconclusive in this case, so we assumed solid body rotation.

The model Stokes~$V$ profiles were fit down to $\rm \chi^2_r=1.6$ (see Fig.~\ref{fig:stokesV}), from an initial value of 7.2. The target $\chi^2_r$ of 1.6 is most likely due to unconstrained differential rotation and slightly underestimated error bars, as the actual noise in the profile looks larger than the uncertainties in some instances. The ZDI magnetic map is illustrated in Fig.~\ref{fig:zdi_maps} and its properties are listed in Table~\ref{tab:zdi_output}. The unsigned, mean magnetic field strength is $\rm B_{mean}=64\,G$, and the topology is predominantly poloidal (64\%), with the dipolar, quadrupolar, and octupolar modes accounting for 67\%, 12\%, and 9\% of the energy. The large-scale topology is also mostly axisymmetric, with 66\% of the energy in the corresponding modes.

\subsection{HD 63433}

For HD~63433, the optimised Voigt kernel parameters are: $d=1.8$, $w_G=1.1\rm\,km\,s^{-1}$, and $w_L=2.6\rm\,km\,s^{-1}$. The stellar inclination was fixed to $70^{\circ}$ and the rotational velocity to 7.3~km~s$^{-1}$ \citep[similar to][]{Bellotti2024}. Our nine Neo-Narval observations in 2024 span 28\,d (or 4.3 rotational cycles), and the differential rotation search yielded a minimum at $\rm P_{rot}=6.489\pm0.075\,d$ and $\rm d\Omega=0.147\pm0.023~rad~d^{-1}$. In a similar manner to the $\rm d\Omega$ result of TOI-1860, the gap in the rotational phases may be introducing biases. The latitudinal differential rotation translates into a rotation period at the pole of $\rm 7.650\pm0.238\,d$ and an equator-pole lap time of 43\,d. The range of $\rm P_{rot}$ values is thus compatible with the rotation period of $\rm 6.45\pm0.05\,d$ given by \citet{Mann2020}.

\begin{figure*}[!t]
    \centering
    \includegraphics[width=0.7\textwidth]{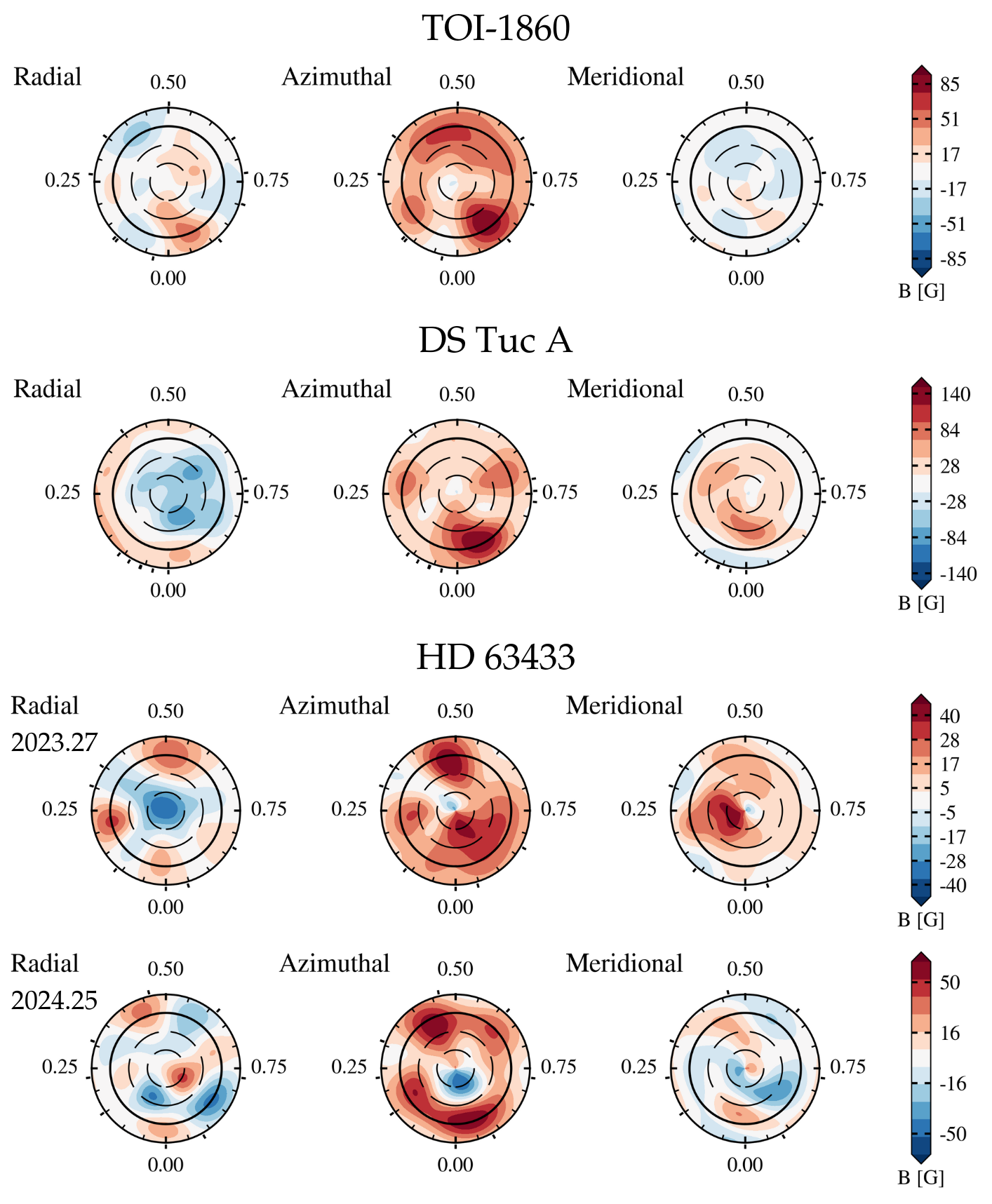}
    \caption{Reconstructed large-scale magnetic field maps of TOI-1860, DS~Tuc~A, and HD~63433 in flattened polar view. For completeness, we included our previous HD~63433 reconstruction using 2023 data \citep{Bellotti2024}. From the left, the radial, azimuthal, and meridional components of the magnetic field vector are illustrated. Concentric circles represent different stellar latitudes: -30\,$^{\circ}$, +30\,$^{\circ}$, and +60\,$^{\circ}$ (dashed lines), as well as the equator (solid line). The radial ticks are located at the rotational phases when the observations were collected. The rotational phases are computed with Eq.~\ref{eq:ephemeris}. The colour bar indicates the polarity and strength (in G) of the magnetic field.}
    \label{fig:zdi_maps}
\end{figure*}

The model Stokes~$V$ profiles were fit down to $\rm \chi^2_r=0.85$ (see Fig.~\ref{fig:stokesV}), from an initial value of 5.45. The ZDI magnetic map is illustrated in Fig.~\ref{fig:zdi_maps} and its properties are listed in Table~\ref{tab:zdi_output}. The mean, unsigned magnetic field strength is $\rm B_{mean}=30\,G$, and the topology features 47\% of the total energy in the poloidal component and 53\% in the toroidal component. The poloidal component shows complexity, with the dipolar, quadrupolar, and octupolar modes storing 7\%, 23\%, and 20\% of the magnetic energy. The large-scale toroidal component is mostly axisymmetric (90\%) while the poloidal component is non-axisymmetric (7\%).

This is the second ZDI reconstruction for HD~63433 after \citep{Bellotti2024}. We can compare our results with the 2023 ZDI reconstruction considering that the cadence and S/N of the observations are similar. We note a difference in the complexity of the large-scale configuration, as shown in Fig.~\ref{fig:zdi_maps} (and reported Table~\ref{tab:zdi_output}). In 2024 the magnetic field is more complex, since it has less energy in the dipolar mode and more in the octupolar and $\ell=4$ modes.

\section{Stellar wind}\label{sec:winds}

We simulated the stellar wind of TOI-1860, DS~Tuc~A, and HD~63433 using the Space Weather Modelling Framework \citep[\texttt{SWMF},\ ][]{Toth2005,Toth2012} and specifically the Alfvén wave solar model \citep[\texttt{AWSoM},\ ][]{Sokolov2013,vanderHolst2014} applied to the ZDI reconstructions described in Sect~\ref{sec:zdi}. A detailed description of the methodology behind the wind models can be found in the recent works of \citet{oFionnagain2019}, \citet{Kavanagh2019}, \citet{Evensberget2021}, \citet{Alvarado-Gomez2022b}, \citet{Evensberget2022}, \citet{BoroSaikia2023} and \citet{Evensberget2023}. Here, we briefly summarise the main features of the model. The three-dimensional simulations of the stellar wind are performed by numerically solving the ideal two-temperature magnetohydrodynamic equations, letting the models converge towards a steady state solution in which balance between the magnetic and hydrodynamic forces is reached across the domain of the simulation \citep[a summary of the equations is given in][]{Evensberget2021}. 

\begin{figure*}[!t]
    \centering
    \includegraphics[width=0.4\textwidth]{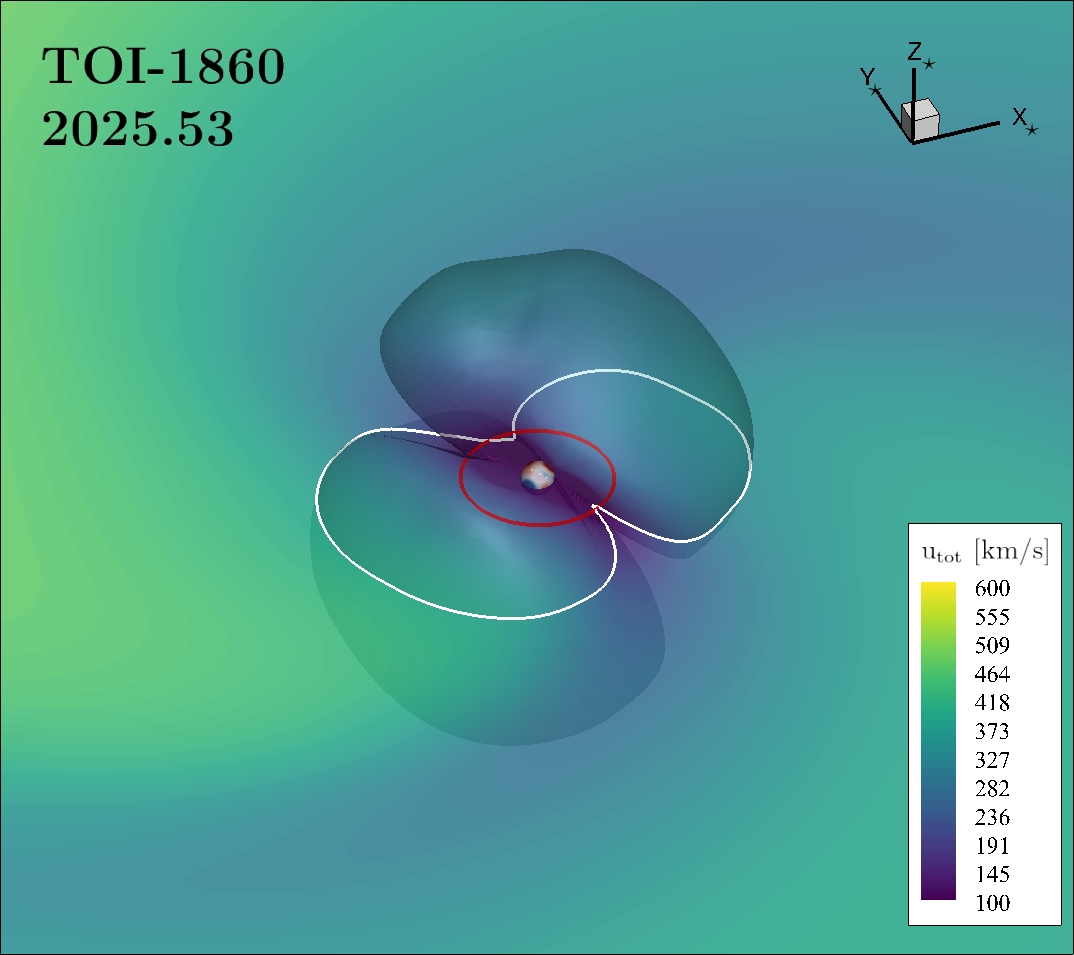}
    \includegraphics[width=0.4\textwidth]{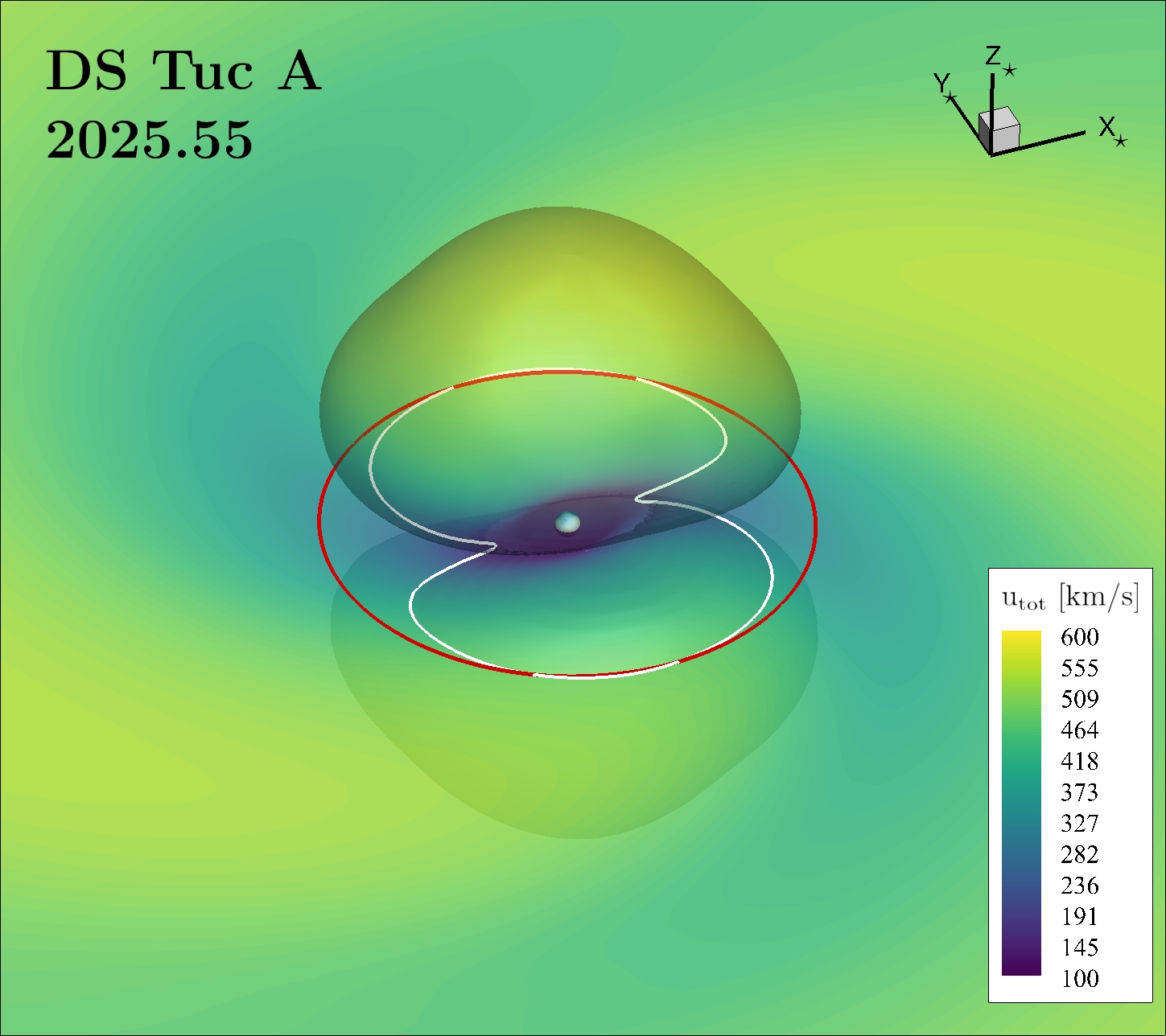}
    \includegraphics[width=0.4\textwidth]{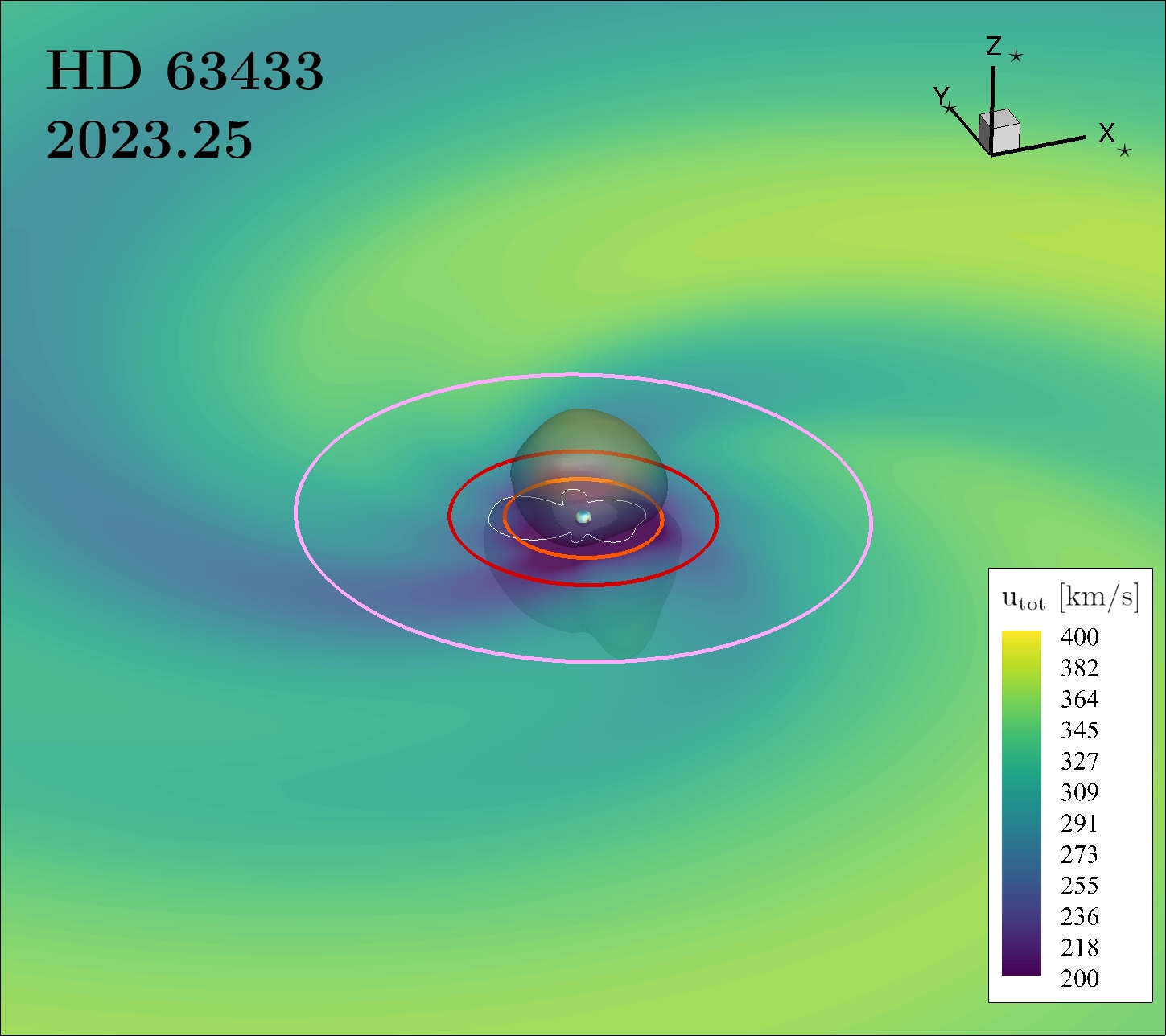}
    \includegraphics[width=0.4\textwidth]{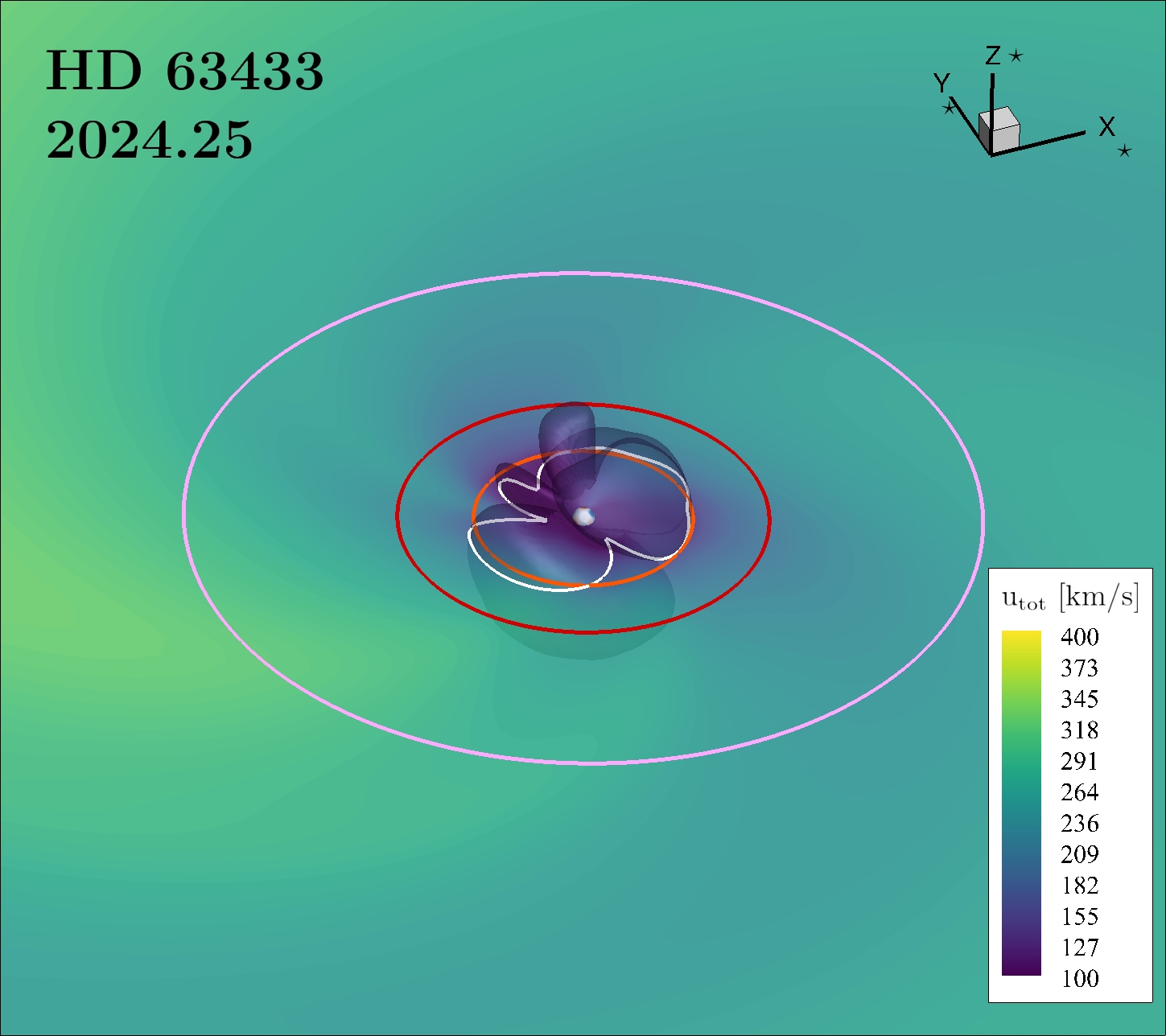}
    \caption{Simulated stellar wind of TOI-1860, DS~Tuc~A, and HD\,63433 in the $\rm x_\star-y_\star$ plane. The rotation axis lies along the positive $\rm z_\star$. The Alfv\'en surface is shown as a translucent surface and its intersection on the $\rm x_\star-y_\star$ plane is shown as a white curve. The colour bar indicates the total wind velocity. The orbits of the hosted exoplanets are also included as coloured ellipses. For HD~63433, we also show the simulations from the 2023 data analysed in \citet{Bellotti2024}.}
    \label{fig:3dwinds}
\end{figure*}

Our \texttt{AWSoM} model extends from the chromosphere (that is, the inner boundary), where the temperature is set at $\rm 5\times10^4~K$ and the number density is set at $\rm 2\times10^{11}~m^{-3}$, through the transition region to the stellar corona. In the \texttt{AWSoM} model, the stellar corona is heated by dissipation of Alfv\'en waves emanating from deeper stellar layers, resulting in a Poynting flux $\rm \Pi_A$ proportional to the local $\rm |\vec{B}|$ value at the inner model boundary. \citet{BoroSaikia2023} used FUV emission lines forming in the chromosphere and the transition region to measure the velocity associated to the non-thermal processes that drive the solar and stellar wind. They found that the non-thermal velocities of Sun-like stars can have similar values as that observed in the solar chromosphere and transition region, hence we expect the energy density of Alfv\'en waves to resemble that of the Sun. The propagation and partial reflection of the Alfv\'en waves results in a turbulent cascade that heats and accelerates the solar wind~\citep{vanderHolst2014, Gombosi2018}. We set the Poynting flux-to-field ratio to $1.1\times10^5$~erg~cm$^{-2}$~G$^{-1}$, the same one as used in solar wind models, and the turbulence correlation length to $1.5\times10^9$~m~T$^{1/2}$.

The models fixes the large-scale radial field component at the inner boundary to the ZDI-derived values (see Fig.~\ref{fig:zdi_maps}), while the transverse components are left to evolve as the numerical solution relaxes towards steady state. We emphasise that, except for the stellar mass, radius, rotation period and large-scale magnetic field, the input parameters of the \texttt{AWSoM} model were set to solar values that have been shown to reproduce solar wind conditions~\citep{Meng2015,vanderHolst2019,Sachdeva2019}. 

\subsection{Characteristics of the wind}

The results of our three-dimensional simulations for TOI-1860, DS~Tuc~A, and HD~63433 are presented in Fig.~\ref{fig:3dwinds}. We also include the simulation of HD~63433 obtained from 2023 data and presented in \citet{Bellotti2024}. Each panel is centred on the star and shows the steady-state solution of the stellar wind, colour-coded by the total wind speed $\rm u_{tot}=\sqrt{u_x^2+u_y^2+u_z^2}$. Moving away from the star, the wind exhibits a spiral shape owing to stellar rotation, and the wind speed (${\bf u}$) increases while the local wind density ($\rm \rho_w$) and magnetic field ($\rm B_w$) decrease, as expected. The panels also illustrate the Alfv\'en surface, which is the boundary where the local wind speed matches the Alfv\'en wave velocity. The latter represents the velocity of magnetic waves propagating through plasma and it is expressed in cgs units as $\rm v_{A} = B_w / \sqrt{4\pi\rho_w}$. As we subsequently discuss in Sect.~\ref{sec:environment}, the Alfv\'en surface is a key factor to determine the possibility of magnetic star-planet interactions \citep[e.g.][]{Vidotto2025}.

For most of our stars, the Alfv\'en surface appears predominantly two-lobed, as expected for stars with dominant dipolar large-scale field configurations \citep[e.g.][]{Evensberget2023}. Consistently with the ZDI reconstructions described in Sect.~\ref{sec:zdi}, the Alfv\'en surface of HD~63433 features a larger and more complex shape than TOI-1860 and DS~Tuc~A, and than our previous wind simulation of HD~63433 corresponding to 2023 data \citep[see][]{Bellotti2024}. This temporal evolution of the Alfv\'en surface, which correlates with the evolution of the large-scale magnetic field, can modulate magnetic star-planet interactions.

We computed the mass loss rate ($\rm \dot{M}$) by integrating the mass flux over a closed spherical surface ($\Sigma$) centred on the star
\begin{equation}\label{eq:mdot}
    \rm \dot{M}=\oint_\Sigma \rho~{\bf u}\cdot\hat{\bf n}~d\Sigma,
\end{equation}
where $\rho$ and ${\bf u} $ are the stellar wind density and speed of our steady-state solution. As listed in Table~\ref{tab:zdi_output}, we estimated values between 5 and 20 times larger than the solar wind-mass loss rate. We note that $\rm \dot{M}$ should not vary with the choice of surface $\Sigma$, as long as such surface encloses the star and the model has reached steady state. We found variations for less than 1\% by changing the spherical surface radius across the simulation domain, as a numerical check to ensure that the simulation has reached steady-state. The values of $\rm \dot{M}$ are reported in Table~\ref{tab:zdi_output}.

We then computed the angular momentum loss rate ($\rm \dot{J}$), which regulates the spin-down of the star with age. Following \citet{Mestel1999} and \citet{Vidotto2014}, 
\begin{equation}\label{eq:jdot}
    \rm \dot{J}=\oint_{\Sigma} \left[ -\frac{\varpi B_\varphi B_r}{4\pi} + \varpi \rho u_\varphi u_r \right] ~d\Sigma,
\end{equation}
where $\rm \varpi = \sqrt{(x^2+y^2)} $ is the cylindrical radius and ($\rm B_r$, $\rm u_r$) and ($\rm B_\varphi$, $\rm u_\varphi$) are the radial and azimuthal components of the magnetic field and speed of the stellar wind. We estimated values between a factor of 1.4 and 35 larger than the average angular momentum loss rate of the Sun computed for cycle 23-24 \citep{Finley2019}, and all conserved at most within 5\%. The values of $\rm \dot{J}$ are reported in Table~\ref{tab:zdi_output}.

\subsection{Environment at the planetary orbit}\label{sec:environment}

\begin{figure*}
    \includegraphics[width=\textwidth]{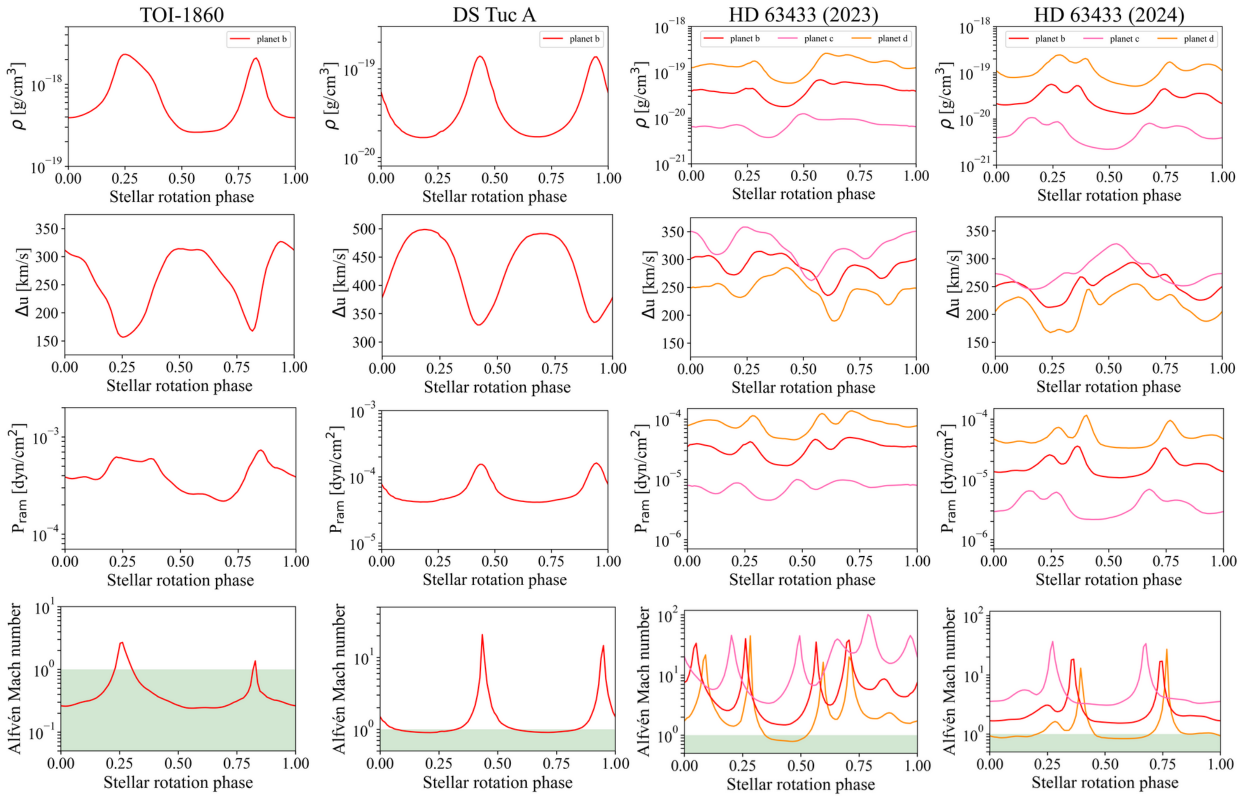}
    \caption{Stellar wind conditions in the planetary frame as a function of stellar rotation phase. From the left, the columns refer to the TOI-1860, DS~Tuc~A, HD~63433 (2023), and HD~63433 (2024) systems. From the top, the panels show the wind density ($\rho$), relative velocity ($\rm\Delta u$), ram pressure ($\rm P_{ram}$), and Alfv\'en Mach number ($\rm M_A$). The sub-Alfv\'enic regime ($\rm M_A<1$) of the stellar wind is shown as a green shaded region. In the $\rm M_A$ panel of HD~63433, we also show the computation for planet d from 2023 data as a dashed line \citep[see][]{Bellotti2024}. }\label{fig:environments}
\end{figure*}

We now describe the stellar wind characteristics at the orbits of the known exoplanets, which are illustrated in Fig.~\ref{fig:3dwinds} as coloured ellipses around the stars. Quantitative information of the wind conditions at these orbits and the location relative to the Alfv\'en surface are shown in Fig.~\ref{fig:environments}. Using the planetary frame as a reference, we computed the variations of wind density, relative speed ($\rm \bf\Delta u$), ram pressure and Alfv\'en Mach number during one stellar rotation. The relative speed is between the wind velocity and the Keplerian velocity of the planet $\rm \bf\Delta u = u - v_K$, and it is used to compute the ram pressure experienced by the planet $\rm P_{ram}=\rho_w\Delta u^2$. The Alfv\'en Mach number is the ratio between the Alfv\'en wave velocity and the wind speed, $\rm M_A = \Delta u / v_A $, and it is a practical quantity to determine whether an orbit is in sub-Alfv\'enic regime $\rm M_A < 1$ or in super-Alfv\'enic regime $\rm M_A > 1$.

TOI-1860 is a young (133\,Myr) K1-type star hosting one known planet at 0.02\,au or 1.066\,d orbit as revealed by $TESS$ observations \citep{Giacalone2022}. The planet radius is $\rm 1.31\,R_\oplus$, the equilibrium temperature is 1885\,K and the authors estimated a mass of $\rm 2.2\,M_\oplus$ from the probabilistic relation of \citet{Chen2017}. The eccentricity of the planetary orbit has not been constrained yet in the literature, hence we assumed a circular orbit in the equatorial plane of the star with radius equal to $\rm 4.66~R_\star$ (with $\rm R_\star = 0.94~R_\odot$). As shown in the first column of Fig.~\ref{fig:environments}, the average wind density and speed are $\rm 7.9\times10^{-19}~g/cm^3$ and $\rm 260\,km/s$, and the planet experiences an average ram pressure of $\rm 4.2\times10^{-4}~dyn/cm^2$. The two dips (or peaks) stem from the spiral morphology of the wind, which is in turn shaped by the change of polarity of the dipolar component. The orbit of TOI-1860\,b is predominantly sub-Alfv\'enic and becomes super-Alfv\'enic briefly between phases 0.23-0.29 and 0.81-0.82. 

DS~Tuc~A is a young (45\,Myr) G5-type star in a visual binary system together with the K3-type star DS~Tuc~B at a separation of 5\,arcsec \citep{Torres2006}. The primary component hosts a $\rm 5.7\,R_\oplus$, super-Neptune planet at an orbital distance of 0.18\,au or period of 8.1\,d \citep{Newton2019,Benatti2019}. By modelling radial velocity observations, \citet{Benatti2019} estimated an upper limit on the exoplanet's mass of $\rm 1.3\,M_{Nep}$ and they considered the planet to be potentially inflated. Recent simulations by \citet{King2025} of future scenarios for the exoplanet showed that, at 5\,Gyr, it may become a Neptune-sized planet or a super-Earth stripped of its primordial \ion{H}/{He} envelope. Assuming a circular orbit in the equatorial plane of the star at $\rm 20.35~R_\star$ \citep{Newton2019} where $\rm R_\star = 0.96~R_\odot$, we found an average wind density, speed, and ram pressure of $\rm 4.2\times10^{-20}~g/cm^3$, $\rm 430\,km/s$, and $\rm 6.5\times10^{-5}~dyn/cm^2$. We note that these values vary of less than 10\% if we assume the orbital distance of $\rm 19.42~R_\star$ obtained by \citet{Benatti2019}. We found that the orbit of the planet is trans-Alfv\'enic, since parts of it are sub-Alfv\'enic (orbital phases between 0.07-0.31 and 0.58-0.83) while the rest is super-Alfv\'enic.

HD~63433 is a young (414\,Myr) G5-type star hosting three planets at 0.05\,au (planet d), 0.07\,au (planet b), 0.14\,au \citep[planet c][]{Capistrant2024}. The orbital periods are 4.2\,d, 7.11\,d, and 20.55\,d, and the radii are 1.07\,R$_\oplus$, 2.02\,R$_\oplus$, and 2.44\,R$_\oplus$ \citep[see also][]{Mann2020,Mallorquin2023,Damasso2023}. From radial velocity observations, \citet{Damasso2023} measured planetary mass upper limits of 11\,M$_\oplus$ for planet b and 31\,M$_\oplus$ for planet c, while \citet{Mallorquin2023} measured 22\,M$_\oplus$ for planet b and 15.5\,M$_\oplus$ for planet c. 

Following \citet{Bellotti2024}, we assumed that the planets have circular orbits in the equatorial plane of the star with radii of $\rm 16.8~R_\star$, $\rm 36.1~R_\star$, and $\rm 9.9~R_\star$ for planet b, c, and d (with $\rm R_\star = 0.897~R_\odot$). For planet b, the average wind density, speed, and ram pressure are $\rm 2.9\times10^{-20}~g/cm^3$, $\rm 250\,km/s$, and $\rm 1.7\times10^{-5}~dyn/cm^2$. For planet c, these values are $\rm 5.3\times10^{-21}~g/cm^3$, $\rm 275\,km/s$, and $\rm 3.9\times10^{-6}~dyn/cm^2$. For planet d, these values are $\rm 1.2\times10^{-19}~g/cm^3$, $\rm 214\,km/s$, and $\rm 5.1\times10^{-5}~dyn/cm^2$.

As for the 2023 observations, the orbits of planet b and c are super-Alf\'enic \citep[see][]{Bellotti2024}. For planet d, the Alfv\'en surface encompasses a larger fraction of the planetary orbit compared to the simulation of 2023 wind conditions (see Fig.~\ref{fig:environments}), owing to the temporal evolution and intensification of the large-scale magnetic field. More specifically, the orbit of planet d was sub-Alfv\'enic between phases 0.38-0.50, while in our new simulation it is between phases 0.00-0.20, 0.45-0.69, and 0.86-1.00 (see the lower right panel of Fig.~\ref{fig:environments}).

\section{Conclusions}\label{sec:conclusions}

With this work, we follow \citet{Bellotti2024} and provide an update on the results of the spectropolarimetric campaign dedicated to characterise the magnetic field of known exoplanet host stars belonging to the current $Ariel$ candidate sample. Our programme aims to collect information on the magnetic activity of a representative sample to inform the observing strategy of $Ariel$. We analysed spectropolarimetric observations of 15 stars obtained with Neo-Narval, HARPSpol, and SPIRou.

We did not detect circularly polarised Zeeman signatures in eight stars over multiple snapshot observations. We detected Zeeman signatures for TOI-1136, TOI-1860, DS~Tuc~A, and the newest observations of HD~63433, and we obtained one marginal detection for TOI-836 and two for TOI-2076. We estimated absolute longitudinal magnetic field values of the order of 5 to 50\,G for these stars, which is in agreement with previous studies of fast-rotating, solar-like stars \citep[e.g.][]{Petit2005,Petit2008,Marsden2014,Folsom2016,Brown2022}. 

We performed the reconstruction of the large-scale magnetic field via ZDI for TOI-1860, DS~Tuc~A, and HD~63433, but not TOI-1136, since we did not have a sufficient number of observations. For TOI-1860 and DS~Tuc~A, this is the first reconstruction of the large-scale magnetic field. For TOI-1860, we found a predominantly toroidal and moderately axisymmetric large-scale magnetic field with an average strength of 37\,G. For DS~Tuc~A, we found a poloidal magnetic field that is mostly dipolar and axisymmetric with an average field strength of 64\,G. For HD~63433, our new 2024 observations revealed a more complex topology compared to the 2023 map, with more energy stored in the $\ell=3$ and $\ell=4$ modes at the expenses of the $\ell=1$ mode. The other features are similar: $\sim$50\% poloidal, moderately axisymmetric and with an average field strength of 70\,G. In a broader context, the ZDI reconstructions of these stars are in agreement with the large-scale field configuration of young, solar-like stars which are known to exhibit variability also of the order of one year \citep[see][for recent examples]{Folsom2016,Willamo2022,Bellotti2025b}.

Finally, we used the magnetic field reconstructions to numerically model the stellar wind environment and compute the location of the Alfv\'en surface. As expected for dipole-dominated, large-scale magnetic field topologies, the shape of the Alfv\'en surface is predominantly two-lobed. For HD~63433, the complexity of the large-scale field, that is the deviation from being simply dipolar, translates in a more complex configuration of the Alfv\'en surface as well. Based on the location of the Alfv\'en surface, we then evaluated the regime in which the hosted exoplanets are orbiting \citep[see][for a recent review]{Vidotto2025}.

We found that the orbit of TOI-1860~b is almost completely sub-Alfv\'enic, meaning that a direct connection between the planet and the stellar magnetic field can occur. As a result, the orbital motion of the planet can generate Alfv\'en waves propagating towards the star or magnetic reconnection events \citep[e.g.][]{Neubauer1998,Ip2004,Saur2013,Strugarek2015,Kavanagh2022}. From the recent work of \citet{Presa2024}, assuming the planetary magnetic field to be dipolar and with a certain obliquity, we expect the planet atmospheric escape in sub-Alfv\'enic regime to occur via a single polar outflow, which was found to be weaker than the bipolar outflow typical of super-Alfv\'enic interactions \citep{Carolan2021b}. We also note that, given the young age of the system, XUV irradiation would also be an important source of atmospheric mass loss.

There are several ways with which the stellar wind can affect atmospheric escape in exoplanets. They can reduce escape rates \citep{Vidotto2020}, change signatures of atmospheric evaporation through spectroscopic transits \citep{Carolan2021}, and possibly generate a tenuous atmosphere through sputtering processes \citep{Vidotto2018}. With the constrained stellar wind properties presented in our work, future atmospheric models of exoplanets will be able to better pin-point the effects of the wind of the host star on the planet’s upper atmosphere, as well as potential signatures in Ly-$\alpha$ and He~\textsc{i} transits. For Ariel specifically, such models are important to assess the presence and evolution of an exoplanet atmosphere \citep{Kubyshkina2022}, as well as to interpret the planetary atmospheric response to stellar activity \citep[e.g. upper-atmosphere heating, ionisation, and chemistry][]{GarciaMunoz2023,Strugarek2025}.

The orbits of HD~63433~b and c are super-Alfv\'enic, in a similar manner as the previous stellar wind model from 2023 data \citep[see][]{Bellotti2024}. In this regime, which is reminiscent of the regime in which Solar System planets orbit, we expect the formation of a bow shock between a potential planetary magnetosphere and the stellar wind \citep{1931TeMAE..36...77C,2009ApJ...703.1734V, Vidotto2010}, as well as the presence of evaporated planetary material in the shape of a tail \citep[e.g.][]{Schneiter2007,Villareal2018}. Finally, planets DS~Tuc~A~b and HD~63433~d are in the trans-Alfvénic region, where the stellar wind conditions can alternate between super- and sub-Alfvénic. For HD~63433~d, we note that sub-Alfv\'enic fraction of the orbit is increased relative to the model from 2023 data, owing to the temporal variability of the stellar large-scale magnetic field. Finally, we also note that the Alfv\'en surface in our model is mostly dictated by the ratio between the Poynting flux and the magnetic field strength reconstructed with ZDI. Such reconstructions may lead to an underestimated magnetic field strength \citep{Lehmann2019}, ultimately affecting the size of the Alfv\'en surface \citep[see e.g. Fig.~5][]{Kavanagh2021}. 

At this point, additional observations are required to confirm whether the large-scale magnetic field is detectable for certain exoplanet host stars. An example of uncertain magnetic field detection is TOI-2076, for which we found two marginal detections and four non-detections, while \citet{Damasso2024} corroborated its magnetic activity and a long-term variation of $\rm\sim2.7\,yr$. Furthermore, our results motivate the spectropolarimetric monitoring of the presented sub-sample of potential $Ariel$ stars on the long-term. Such a monitoring can address the question of whether the stars manifest magnetic cycles and if so, how they relate to the activity cycles discovered with other techniques. For instance, DS~Tuc~A exhibits a long-term photometric variation of $\rm\sim8\,yr$, as described by \citet{Benatti2019}. Ultimately, this will be valuable to constrain the temporal evolution of the Alfv\'en surface and of the observational signatures marking magnetic star-planet interactions, as well as the evolution of evaporating atmospheres and their interactions with the stellar wind.

The spectropolarimetric programme presented here started in 2022 and it is expected to continue until the launch of $Ariel$, with the aim of informing observing strategies for specific targets as well as atmospheric modelling. Once the mission is launched, the goal of the campaign will become of a follow-up nature, thus complementing the observations already performed and providing new insights for the interpretation of $Ariel$ data. The $Ariel$ Candidate Sample is subject to yearly adjustments based on new analyses of the most suitable targets for atmospheric characterisation, but also in light of new exoplanet discoveries. Although it is hard to estimate a number of targets given the dynamic nature of the Candidate Sample combined with our selection criteria for magnetically active stars, we expect between two and five additional stars per year, for which to assess the magnetic field detectability and reconstruct the topology.

\section{Data Availability}

The spectropolarimetric observations analysed in this work are available on online databases. All Neo-Narval and SPIRou observations are available on PolarBase\footnote{\url{https://www.polarbase.ovgso.fr/}} \citep{Petit2014}. The Neo-Narval observations were taken under the programmes L232N02, L241N09, and L251N08 and the SPIRou observations under the programmes 22BF97, 23AF16, 23BF98, and 24AF99. The HARPSPol observations are available at the ESO Science Archive\footnote{See \url{https://archive.eso.org/wdb/wdb/eso/repro/form}} under the programmes 110.24C8.001, 110.24C8.002, 115.28DD.001, and 115.28DD.002. 

\begin{acknowledgements}

This publication is part of the project "Exo-space weather and contemporaneous signatures of star-planet interactions" (with project number OCENW.M.22.215 of the research programme "Open Competition Domain Science- M"), which is financed by the Dutch Research Council (NWO). This work used the Dutch national e-infrastructure with the support of the SURF Cooperative using grant nos. EINF-2218 and EINF-5173. This work has been developed within the framework of the Ariel ``Stellar Characterisation'' and ``Stellar Activity'' working groups, in synergy with the ``Planetary Formation'' working group of the ESA $Ariel$ space mission Consortium. AAV and DE acknowledge funding from the European Research Council (ERC) under the European Union's Horizon 2020 research and innovation programme (grant agreement No 817540, ASTROFLOW). AAV acknowledges funding from the Dutch Research Council (NWO), with project number VI.C.232.041 of the Talent Programme Vici.
We thank the TBL team for providing service observing with Neo-Narval under the programmes  L232N02, L241N09, and L251N08 (PI S. Bellotti). Based on observations obtained at the Canada-France-Hawaii Telescope (CFHT) which is operated by the National Research Council of Canada, the Institut National des Sciences de l'Univers of the Centre National de la Recherche Scientique of France, and the University of Hawaii. The observations were collected under the programmes 22BF97, 23AF16, 23BF98, and 24AF99 (PI S. Bellotti and P. Petit). Based on observations collected at the European Southern Observatory under ESO programmes 110.24C8.001, 110.24C8.002, 115.28DD.001, and 115.28DD.002 (PI. S. Bellotti). 
This research has made use of the NASA Exoplanet Archive, which is operated by the California Institute of Technology, under contract with the National Aeronautics and Space Administration under the Exoplanet Exploration Program. This work made use of the Ariel Stellar Catalogue developed by the Stellar Characterisation WG in preparation of the ESA Ariel space mission. This work used the BATS-R-US tools developed at the University of Michigan Center for Space Environment Modeling and made available through the NASA Community Coordinated Modeling Center. This work has made use of the VALD database, operated at Uppsala University, the Institute of Astronomy RAS in Moscow, and the University of Vienna; Astropy, 12 a community-developed core Python package for Astronomy \citep{Astropy2013,Astropy2018}; NumPy \citep{VanderWalt2011}; Matplotlib: Visualization with Python \citep{Hunter2007}; SciPy \citep{Virtanen2020} and PyAstronomy \citep{Czesla2019}.

\end{acknowledgements}

%
%

\bibliographystyle{aa}
\bibliography{biblio}

\begin{appendix}

\onecolumn

\section{Additional figures}\label{sec:stokesV}

\begin{figure}[h]
\centering
    \includegraphics[width=\textwidth]{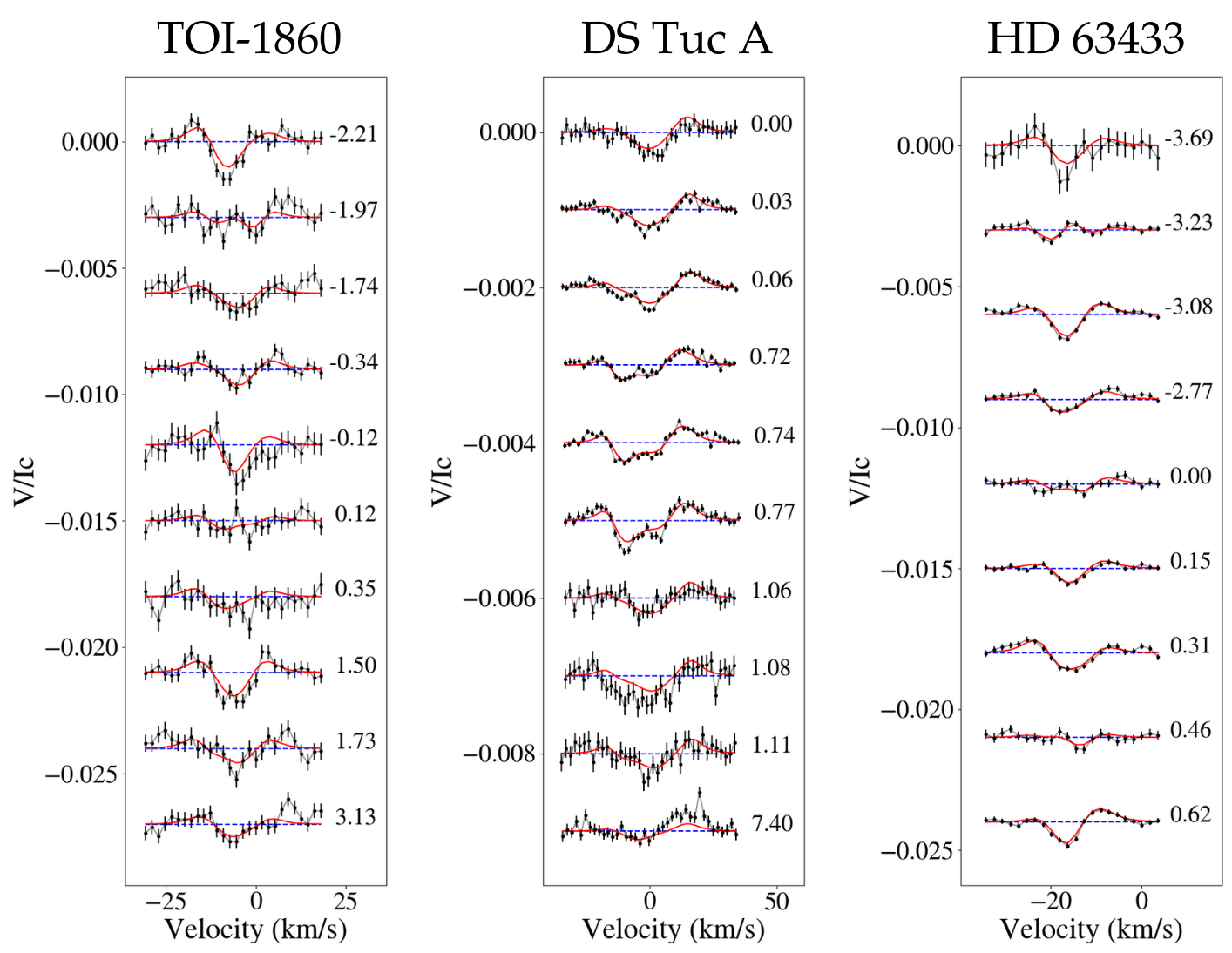}
    \caption{Time series of Stokes~$V$ LSD profiles and the ZDI models for TOI-1860, DS~Tuc~A, and HD~63433. The observations are shown in black and the models in red. The numbers on the right indicate the rotational cycle computed from Eq.~\ref{eq:ephemeris} using the first observation of an epoch as reference date for DS~Tuc~A and the median observation for TOI-1860 and HD~63433. The horizontal line represents the zero point of the profiles, which are shifted vertically based on their rotational phase for visualisation purposes.}
    \label{fig:stokesV}
\end{figure}

In this appendix we provide the observed Stokes~$V$ LSD profiles used for the ZDI reconstruction and their Unno-Rachkovsky models. Fig.~\ref{fig:stokesV} shows the Stokes~$V$ LSD profiles for TOI-1860, DS~Tuc~A, and HD~63433 analysed in this work.

\begin{landscape}

\section{Journal of observations}\label{app:log}

In this appendix, we provide a list of the observations performed with Neo-Narval, HARPSpol, and SPIRou. We also list the value of the longitudinal magnetic field ($\rm B_\ell$).
\setlength{\tabcolsep}{5pt}
\begin{longtable}{lrccrrrccrrccr}
\caption{Journal of observations}\label{tab:log}\\
\toprule
Star & Instrument & Date & UT & HJD & $\rm t_{exp}$ & S/N  & $\rm FAP_V$ & $\rm Detection_V$ & $\rm\sigma_{LSD,V}$ & $\rm B_\ell$ & $\rm FAP_N$ & $\rm Detection_N$ & $\rm \sigma_{LSD,N}$\\
 & & [yyyy-mm-dd] & [hh:mm:ss] & [$-2450000$] & [s] & &  & & [$10^{-4}I_c$] & [G] & & & [$10^{-4}I_c$] \\
\midrule
\endfirsthead
\caption{continued.}\\
\toprule
Star & Instrument & Date & UT & HJD & $\rm t_{exp}$ & S/N  & $\rm FAP_V$ & $\rm Detection_V$ & $\rm\sigma_{LSD,V}$ & $\rm B_\ell$ & $\rm FAP_N$ & $\rm Detection_N$ & $\rm \sigma_{LSD,N}$\\
 & & [yyyy-mm-dd] & [hh:mm:ss] & [$-2450000$] & [s] & & & & [$10^{-4}I_c$] & [G] & & & [$10^{-4}I_c$] \\
\midrule
\endhead
\bottomrule
\endfoot
HD~63433 & Neo-Narval & 2024-03-18 & 20:51:40.58 & 10388.3679 & 4x883 & 326 & 1.80E-01 & ND & 2.41 & $-0.9\pm8.0$ & 9.94E-01 & ND & 2.7\\
HD~63433 & Neo-Narval & 2024-03-21 & 20:54:30.59 & 10391.3695 & 4x883 & 505 & 1.52E-06 & D  & 1.10 & $-0.9\pm2.1$ & 6.52E-01 & ND & 1.03\\
HD~63433 & Neo-Narval & 2024-03-22 & 20:03:28.64 & 10392.3340 & 4x883 & 605 & 0.00E-00 & D  & 1.25 & $ 0.0\pm1.6$ & 9.76E-01 & ND & 0.85\\
HD~63433 & Neo-Narval & 2024-03-24 & 20:44:58.99 & 10394.3626 & 4x883 & 527 & 0.00E-00 & D  & 1.09 & $-4.4\pm1.9$ & 6.33E-01 & ND & 0.98\\
HD~63433 & Neo-Narval & 2024-04-11 & 19:58:43.77 & 10412.3290 & 4x883 & 408 & 4.86E-02 & ND & 1.37 & $-2.9\pm2.6$ & 9.47E-01 & ND & 1.27\\
HD~63433 & Neo-Narval & 2024-04-12 & 19:33:13.52 & 10413.3112 & 4x883 & 578 & 0.00E-00 & D  & 0.98 & $-3.1\pm1.7$ & 4.14E-02 & ND & 0.88\\
HD~63433 & Neo-Narval & 2024-04-13 & 19:48:49.81 & 10414.3219 & 4x883 & 488 & 0.00E-00 & D  & 1.26 & $ 5.0\pm1.9$ & 2.11E-01 & ND & 1.00\\
HD~63433 & Neo-Narval & 2024-04-14 & 20:21:20.89 & 10415.3444 & 4x883 & 396 & 2.43E-02 & ND & 1.32 & $ 3.3\pm2.8$ & 6.67E-01 & ND & 1.39\\
HD~63433 & Neo-Narval & 2024-04-15 & 20:01:09.13 & 10416.3304 & 4x883 & 552 & 0.00E-00 & D  & 1.19 & $-5.7\pm1.6$ & 9.92E-01 & ND & 0.80\\
HD~63935  &  Neo-Narval  & 2023-11-19 & 03:28:06.76 & 10267.6497 &  4x898  & 243 & 1.00E-01 &  ND  & 2.81 &  $<30$  & 3.70E-01 &  ND  &  2.87\\
HD~63935  &  Neo-Narval  & 2023-11-20 & 03:19:59.39 & 10268.6441 &  4x898  & 234 & 6.53E-01 &  ND  & 2.48 &  $<32$  & 6.48E-01 &  ND  &  2.70\\
HD~63935  &  Neo-Narval  & 2023-11-26 & 03:30:11.50 & 10274.6514 &  4x898  & 243 & 1.00E+00 &  ND  & 3.15 &  $<50$  & 9.84E-01 &  ND  &  3.27\\
HD~63935  &  Neo-Narval  & 2024-01-10 & 00:23:49.40 & 10319.5213 &  4x898  & 243 & 6.82E-01 &  ND  & 2.39 &  $<30$  & 4.13E-01 &  ND  &  2.56\\
HD~63935  &  HARPSpol  & 2022-12-24 & 05:05:10.35 & 9937.7169 &  4x900  & 295 & 9.38E-01 &  ND  & 1.68 &  $<20$  & 5.34E-01 &  ND  &  1.71\\
HD~63935  &  HARPSpol  & 2022-12-26 & 04:27:45.35 & 9939.6910 &  4x900  & 190 & 9.76E-02 &  ND  & 2.22 &  $<20$  & 6.35E-02 &  ND  &  2.07\\
HD~63935  &  HARPSpol  & 2023-02-06 & 04:54:38.00 & 9981.7099 &  4x900  & 220 & 6.90E-01 &  ND  & 1.37 &  $<270$  & 2.56E-01 &  ND  &  1.53\\
HD~63935  &  HARPSpol  & 2023-02-07 & 03:36:16.32 & 9982.6554 &  4x900  & 310 & 9.99E-01 &  ND  & 1.71 &  $<110$  & 9.99E-01 &  ND  &  1.67\\
HD~89345  &  Neo-Narval  & 2024-01-10 & 03:37:48.25 & 10319.6570 &  4x868  & 5 & 9.56E-01 &  ND  & 49.72 &  $<800$  & 9.88E-01 &  ND  &  48.74\\
HD~89345  &  Neo-Narval  & 2024-01-13 & 03:48:17.25 & 10322.6642 &  4x868  & 152 & 8.77E-01 &  ND  & 2.75 &  $<30$  & 7.05E-01 &  ND  &  2.58\\
HD~89345  &  Neo-Narval  & 2024-01-20 & 02:45:43.64 & 10329.6208 &  4x868  & 73 & 4.59E-01 &  ND  & 8.36 &  $<120$  & 5.28E-01 &  ND  &  8.15\\
HD~89345  &  Neo-Narval  & 2024-01-27 & 02:17:24.19 & 10336.6010 &  4x868  & 56 & 2.41E-01 &  ND  & 8.21 &  $<100$  & 2.75E-02 &  ND  &  8.26\\
HD~89345  &  Neo-Narval  & 2024-02-18 & 02:44:54.32 & 10358.6192 &  4x868  & 45 & 3.07E-02 &  ND  & 17.94 &  $<230$  & 4.71E-03 &  MD  &  16.81\\
HD~152843  &  Neo-Narval  & 2023-09-24 & 18:46:43.76 & 10212.2787 &  4x846  & 111 & 4.11E-01 &  ND  & 5.74 &  $<90$  & 9.45E-01 &  ND  &  4.64\\ 
HD~152843  &  Neo-Narval  & 2023-10-02 & 19:06:31.76 & 10220.2923 &  4x846  & 21 & 2.91E-01 &  ND  & 24.73 &  $<350$  & 1.60E-02 &  ND  &  25.67\\ 
HD~152843  &  Neo-Narval  & 2023-10-03 & 18:45:57.52 & 10221.2779 &  4x846  & 143 & 9.47E-01 &  ND  & 3.65 &  $<50$  & 7.03E-01 &  ND  &  3.57\\
HD~158259  &  Neo-Narval  & 2023-09-05 & 19:15:13.11 & 10193.3012 &  4x893  & 413 & 6.81E-01 &  ND  & 1.49 &  $<9$  & 1.44E-02 &  ND  &  1.39\\
HD~158259  &  Neo-Narval  & 2023-09-10 & 19:40:16.58 & 10198.3186 &  4x893  & 689 & 1.53E-01 &  ND  & 0.78 &  $<5$  & 9.98E-01 &  ND  &  0.84\\
HD~158259  &  Neo-Narval  & 2023-09-23 & 19:13:46.56 & 10211.3000 &  4x893  & 482 & 4.16E-01 &  ND  & 1.14 &  $<7$  & 9.43E-01 &  ND  &  1.24\\
HD~260655   &  SPIRou  & 2024-03-03 & 05:06:56.37 & 10372.7118 &  4x1203  & 222 & 9.29E-01 &  ND  & 2.81 &  $<130$  & 8.98E-01 &  ND  &  2.32\\
HAT-P-22  &  Neo-Narval  & 2024-03-12 & 23:12:19.17 & 10382.4681 &  4x1207  & 45 & 8.47E-01 &  ND  & 8.75 &  $<50$  & 2.28E-01 &  ND  &  9.76\\
HAT-P-22  &  Neo-Narval  & 2024-03-13 & 23:22:33.80 & 10383.4751 &  4x1207  & 102 & 5.16E-02 &  ND  & 10.16 &  $<50$  & 4.11E-01 &  ND  &  9.70\\
DSTuc & HARPSpol & 2025-07-18 & 05:01:41.36 & 10874.7477 & 4x1499 & 385 & 1.66E-04 &  MD  & 0.63 &  $-11.4\pm8.7$  & 1.00E-00 &  ND  &  0.48\\
DSTuc & HARPSpol & 2025-07-18 & 07:14:14.59 & 10874.8399 & 4x1499 & 437 & 0.00E-00 &  D  & 0.59 &  $-16.9\pm4.1$  & 9.71E-01 &  ND  &  0.38\\
DSTuc & HARPSpol & 2025-07-18 & 08:57:30.59 & 10874.9116 & 4x1499 & 372 & 0.00E-00 &  D  & 0.61 &  $-19.4\pm3.5$  & 7.57E-01 &  ND  &  0.38\\
DSTuc & HARPSpol & 2025-07-20 & 06:10:29.01 & 10876.7955 & 4x1499 & 359 & 0.00E-00 &  D  & 0.53 &  $-21.3\pm3.6$  & 7.80E-01 &  ND  &  0.37\\
DSTuc & HARPSpol & 2025-07-20 & 07:53:27.62 & 10876.8670 & 4x1499 & 419 & 0.00E-00 &  D  & 0.63 &  $-29.1\pm3.2$  & 3.73E-01 &  ND  &  0.39\\
DSTuc & HARPSpol & 2025-07-20 & 09:36:14.21 & 10876.9314 & 4x1199 & 351 & 0.00E-00 &  D  & 0.75 &  $-21.8\pm5.0$  & 9.96E-01 &  ND  &  0.42\\
DSTuc & HARPSpol & 2025-07-21 & 05:28:24.27 & 10877.7663 & 4x1499 & 164 & 8.30E-02 &  ND  & 0.95 &  $-11.4\pm9.1$  & 4.02E-01 &  ND  &  0.90\\
DSTuc & HARPSpol & 2025-07-21 & 07:11:36.07 & 10877.8379 & 4x1499 & 232 & 2.77E-05 &  D  & 1.10 &  $-21.6\pm11.7$  & 5.20E-01 &  ND  &  0.96\\
DSTuc & HARPSpol & 2025-07-21 & 08:54:19.49 & 10877.9093 & 4x1499 & 162 & 4.07E-02 &  ND  & 1.00 &  $-4.4\pm11.7$  & 9.27E-01 &  ND  &  0.92\\
DSTuc & HARPSpol & 2025-08-08 & 07:05:44.15 & 10895.8333 & 4x1499 & 349 & 1.68E-13 &  D  & 0.79 &  $-36.3\pm6.8$  & 6.71E-03 &  MD  &  0.67\\
Kepler-444  &  Neo-Narval  & 2023-09-20 & 20:24:00.13 & 10208.3498 &  4x857  & 166 & 3.85E-01 &  ND  & 2.19 &  $<40$  & 9.84E-01 &  ND  &  2.07\\
Kepler-444  &  Neo-Narval  & 2023-09-23 & 20:20:22.85 & 10211.3472 &  4x857  & 162 & 9.25E-01 &  ND  & 2.43 &  $<43$  & 5.29E-01 &  ND  &  2.50\\
Kepler-444  &  Neo-Narval  & 2023-09-24 & 20:24:17.13 & 10212.3499 &  4x857  & 154 & 5.26E-01 &  ND  & 2.77 &  $<60$  & 5.31E-01 &  ND  &  2.55\\
Kepler-444  &  Neo-Narval  & 2023-10-02 & 20:19:47.91 & 10220.3464 &  4x857  & 100 & 5.33E-01 &  ND  & 5.92 &  $<110$  & 6.09E-01 &  ND  &  5.28\\
K2-116      &  SPIRou  & 2023-07-01 & 10:38:27.93 & 10126.9490 &  4x902  & 285 & 2.52E-01 &  ND  & 0.94 &  $<12$  & 1.03E-01 &  ND  &  0.96\\
K2-116      &  SPIRou  & 2023-07-26 & 14:20:44.47 & 10152.1035 &  4x902  & 272 & 1.60E-01 &  ND  & 1.07 &  $<14$  & 2.52E-01 &  ND  &  1.04\\
K2-116      &  SPIRou  & 2023-10-03 & 05:00:18.87 & 10220.7100 &  4x902  & 235 & 9.90E-01 &  ND  & 1.14 &  $<13$  & 7.05E-01 &  ND  &  1.26\\
TOI-836     &  SPIRou  & 2024-03-15 & 12:45:53.35 & 10385.0374 &  4x1203  & 76 & 9.30E-03 &  MD  & 27.59 &  $62\pm190$  & 1.22E-05 &  D  &  33.57\\
TOI-1136  &  Neo-Narval  & 2024-03-13 & 00:38:36.67 & 10382.5280 &  4x896  & 108 & 2.76E-01 &  ND  & 3.27 &  $<42$  & 2.89E-01 &  ND  &  3.32\\
TOI-1136  &  Neo-Narval  & 2024-03-14 & 00:52:55.94 & 10383.5379 &  4x896  & 79 & 1.13E-01 &  ND  & 11.93 &  $<70$  & 2.00E-01 &  ND  &  10.79\\
TOI-1136  &  Neo-Narval  & 2024-03-16 & 23:28:46.44 & 10386.4793 &  4x896  & 17 & 1.68E-04 &  MD  & 54.71 &  $120\pm106$  & 2.87E-07 &  D  &  59.52\\
TOI-1136  &  Neo-Narval  & 2024-03-18 & 23:03:40.12 & 10388.4618 &  4x896  & 22 & 1.52E-04 &  MD  & 36.69 &  $25\pm44$  & 6.73E-04 &  MD  &  34.77\\
TOI-1136  &  Neo-Narval  & 2024-03-19 & 21:26:21.83 & 10389.3942 &  4x896  & 17 & 1.63E-05 &  D   & 49.39 &  $131\pm60$  & 1.11E-03 &  MD  &  45.71\\
TOI-1136  &  Neo-Narval  & 2024-03-21 & 22:49:35.78 & 10391.4519 &  4x896  & 103 & 3.38E-01 &  ND  & 3.68 &  $<52$  & 1.19E-01 &  ND  &  3.72\\
TOI-1860  &  Neo-Narval  & 2024-03-25 & 01:18:49.96 & 10394.5557 &  4x881  & 91 & 3.31E-02 &  ND  & 3.38 &  $-0.4\pm8.1$  & 2.81E-02 &   ND    &  3.54\\
TOI-1860  &  Neo-Narval  & 2024-04-12 & 00:11:05.63 & 10412.5082 &  4x881  & 139 & 9.04E-03 &  MD  & 2.32 &  $-2.0\pm5.8$  & 2.92E-02 &   ND   &  2.32\\
TOI-1860  &  Neo-Narval  & 2024-04-14 & 00:59:00.79 & 10414.5414 &  4x881  & 222 & 3.79E-02 &  ND  & 3.81 &  $10.3\pm10.4$  & 7.73E-01 &   ND    &  3.81\\
TOI-1860  &  Neo-Narval  & 2024-04-16 & 01:48:11.36 & 10416.5755 &  4x881  & 256 & 1.51E-03 &  MD  & 5.57 &  $6.9\pm12.9$  & 9.87E-01 &   ND    &  5.14\\
TOI-1860  &  Neo-Narval  & 2024-08-09 & 20:15:41.51 & 10532.3425 &  4x881  & 186 & 4.06E-01 &  ND  & 3.03 &  $-3.0\pm8.2$  & 3.05E-01 &   ND    &  3.32\\
TOI-1860  &  Neo-Narval  & 2024-08-15 & 20:10:07.52 & 10538.3386 &  4x881  & 197 & 4.63E-01 &  ND  & 3.75 &  $4.6\pm9.7$  & 4.73E-01 &   ND   &  3.84\\
TOI-1860  &  Neo-Narval  & 2025-06-29 & 21:34:18.71 & 10856.3973 &  4x881  & 268 & 1.98E-03 &  MD  & 5.28 &  $18.7\pm14.4$  & 4.37E-01 &   ND  &  4.98\\
TOI-1860  &  Neo-Narval  & 2025-06-30 & 21:54:41.27 & 10857.4114 &  4x881  & 229 & 2.24E-13 &  D   & 3.30 &  $-4.5\pm6.6$  & 1.86E-01 &   ND   &  2.66\\
TOI-1860  &  Neo-Narval  & 2025-07-01 & 22:09:10.24 & 10858.4215 &  4x881  & 218 & 5.19E-09 &  D   & 3.00 &  $10.4\pm6.5$  & 1.23E-02 &   ND   &  2.46\\
TOI-1860  &  Neo-Narval  & 2025-07-07 & 22:20:38.77 & 10864.4293 &  4x881  & 294 & 4.45E-02 &  ND  & 3.66 &  $-8.5\pm8.5$  & 4.74E-01 &   ND   &  3.51\\
TOI-1860  &  Neo-Narval  & 2025-07-15 & 21:36:13.20 & 10872.3984 &  4x881  & 242 & 4.81E-01 &  ND  & 12.41 &  $31.3\pm31.3$  & 7.62E-01 &   ND   &  11.32\\
TOI-1860  &  Neo-Narval  & 2025-07-16 & 21:30:44.22 & 10873.3946 &  4x881  & 218 & 7.77E-16 &  D   & 2.95 &  $-2.8\pm6.4$  & 4.35E-01 &   ND    &  2.43\\
TOI-1860  &  Neo-Narval  & 2025-07-22 & 21:40:37.13 & 10879.4014 &  4x881  & 278 & 1.66E-02 &  ND  & 3.10 &  $-13.0\pm8.3$  & 9.92E-01 &   ND    &  2.76\\
TOI-1860  &  Neo-Narval  & 2025-07-08 & 22:05:04.57 & 10865.4185 &  4x881  & 188 & 2.89E-12 &  D   & 3.05 &  $2.3\pm6.6$  & 6.62E-01 &   ND  &  2.48\\
TOI-1860  &  Neo-Narval  & 2025-07-09 & 21:57:32.90 & 10866.4133 &  4x881  & 233 & 5.71E-03 &  MD  & 3.12 &  $19.4\pm8.0$  & 9.32E-01 &   ND   &  3.19\\
TOI-1860  &  Neo-Narval  & 2025-07-10 & 22:10:26.97 & 10867.4222 &  4x881  & 224 & 1.29E-03 &  MD  & 2.70 &  $-20.5\pm6.4$  & 5.78E-01 &   ND    &  2.52\\
TOI-1860  &  Neo-Narval  & 2025-08-15 & 20:01:16.85 & 10903.3325 & 4x881 & 115 & 3.76E-05 &  D   & 5.82 &  $-8.5\pm8.0$  & 9.76E-01 &   ND  &  4.93\\
TOI-1860  &  Neo-Narval  & 2025-08-16 & 20:00:36.42 & 10904.3320 & 4x881 & 133 & 5.29E-01 &  ND  & 2.75 &  $-43.5\pm19.3$  & 8.78E-01 &   ND   &  2.59\\
TOI-1860  &  Neo-Narval  & 2025-08-17 & 20:00:12.42 & 10905.3318 & 4x881 & 140 & 4.27E-01 &  ND  & 2.49 &  $-19.8\pm15.4$  & 2.31E-01 &   ND    &  2.43\\
TOI-1860  &  Neo-Narval  & 2025-08-23 & 19:54:42.98 & 10911.3280 & 4x881 & 203 & 8.95e-01 &  ND  & 1.33 & $-4.4\pm9.2$ & 6.44e-01 &  ND  & 1.32\\
TOI-2076  &  Neo-Narval  & 2024-01-20 & 05:31:37.14 & 10329.7332 &  4x900  & 97 & 2.45E-01 &  ND  & 4.70 &  $<98$  & 9.97E-01 &  ND  &  4.56\\
TOI-2076  &  Neo-Narval  & 2024-01-26 & 05:01:18.48 & 10335.7124 &  4x900  & 160 & 2.90E-02 &  MD  & 2.67 &  $-75\pm17$  & 4.49E-01 &  ND  &  3.45\\
TOI-2076  &  Neo-Narval  & 2024-02-18 & 04:48:15.35 & 10358.7038 &  4x900  & 101 & 1.73E-03 &  ND  & 6.40 &  $<102$  & 9.45E-01 &  ND  &  6.02\\
TOI-2076  &  Neo-Narval  & 2024-03-22 & 01:51:51.71 & 10391.5810 &  4x851  & 104 & 7.55E-03 &  MD  & 6.84 &  $69\pm34$  & 5.07E-02 &  ND  &  6.59\\
TOI-2076  &  Neo-Narval  & 2024-03-25 & 00:16:05.50 & 10394.5144 &  4x851  & 85 & 6.20E-01 &  ND  & 5.31 &  $<150$  & 9.55E-01 &  ND  &  5.63\\
TOI-2076  &  Neo-Narval  & 2024-04-11 & 23:06:52.55 & 10412.4657 &  4x851  & 87 & 3.12E-01 &  ND  & 8.80 &  $<93$  & 5.83E-01 &  ND  &  9.08\\
Wolf~503    &  SPIRou  & 2023-07-24 & 06:13:55.10 & 10149.7558 &  4x902  & 324 & 7.27E-01 &  ND  & 37.13 &  $<25$  & 8.50E-01 &  ND  &  42.32\\
Wolf~503    &  SPIRou  & 2024-03-01 & 13:15:28.64 & 10371.0582 &  4x1203  & 156 & 8.78E-01 &  ND  & 0.51 &  $<190$  & 9.96E-01 &  ND  &  0.48\\
Wolf~503    &  SPIRou  & 2024-03-15 & 11:06:41.78 & 10384.9687 &  4x1203  & 61 & 3.68E-01 &  ND  & 5.71 &  $<500$  & 3.19E-01 &  ND  &  5.68\\
\end{longtable}

\justifying\noindent The following quantities are listed: star identifier, instrument used for the observations, date and time of the observation, Heliocentric Julian date, exposure time per polarimetric sequence, S/N per polarimetric sequence at 650\,nm (for Neo-Narval and HARPSpol) or at 1650\,nm (for SPIRou) per resolution pixel, false-alarm probability for Stokes~$V$ detection, status of Stokes~$V$ detection (D = detection, MD = marginal detection and ND = non-detection), RMS noise level of Stokes~$V$ signal in units of unpolarised continuum, longitudinal magnetic field or $3\sigma$ upper limit computed using Eq.~\ref{eq:Bl}, false-alarm probability for Stokes~$N$ detection, status of Stokes~$N$ detection, and RMS noise level of Stokes~$N$ signal in units of unpolarised continuum.

\end{landscape}

\FloatBarrier 
\clearpage
  
\end{appendix}

\end{document}